\documentclass[showpacs,prl,onecolumn,aps,superscriptaddress,preprintnumbers,nofootinbib]{revtex4}
\usepackage[T1]{fontenc}
\usepackage[latin9]{inputenc}
\usepackage{amsmath,amssymb,bigints}
\usepackage{epsfig}
\usepackage{graphicx}
\usepackage{amsmath}
\usepackage{amsfonts}
\def\slashchar#1{\setbox0=\hbox{$#1$}     		
   \dimen0=\wd0                                 	
   \setbox1=\hbox{/} \dimen1=\wd1               	
   \ifdim\dimen0>\dimen1                        	
      \rlap{\hbox to \dimen0{\hfil/\hfil}}      	
      #1                                        	
   \else                                        	
      \rlap{\hbox to \dimen1{\hfil$#1$\hfil}}   	
      /                                         	
   \fi}

\renewcommand{\vec}{\boldsymbol}
\newcommand{\beq}{\begin{equation}}
\newcommand{\eeq}{\end{equation}}
\newcommand{\bea}{\begin{eqnarray}}
\newcommand{\eea}{\end{eqnarray}}
\newcommand{\baa}{\begin{array}}
\newcommand{\eaa}{\end{array}}

\def\eq#1{{Eq.~(\ref{#1})}}
\def\fig#1{{Fig.~\ref{#1}}}

\newcommand{\bas}{\bar{\alpha}_S}
\newcommand{\as}{\alpha_S}

\newcommand{\nn}{\nonumber}

\newcommand{\h}{\frac{1}{2}}

\newcommand{\kv}{\vec{k}}

\newcommand{\pv}{\vec{p}}

\newcommand{\rv}{\vec{r}}
\newcommand{\bv}{\vec{b}}

\newcommand{\Lb}{\left(}
\newcommand{\Rb}{\right)}
\def\pom{{I\!\!P}}

\renewcommand{\vec}[1]{\boldsymbol{#1}}

\def\pom{{I\!\!P}}

\begin{document}
\title{ Energy evolution of  J/$\mathbf{\psi}$ production in DIS on nuclei}
\author{ E.~ Gotsman,}
\email{gotsman@post.tau.ac.il}
\affiliation{Department of Particle Physics, School of Physics and Astronomy,
Raymond and Beverly Sackler
 Faculty of Exact Science, Tel Aviv University, Tel Aviv, 69978, Israel}
 \author{ E.~ Levin}
 \email{leving@post.tau.ac.il, eugeny.levin@usm.cl} \affiliation{Department of Particle Physics, School of Physics and Astronomy,
Raymond and Beverly Sackler
 Faculty of Exact Science, Tel Aviv University, Tel Aviv, 69978, Israel} 
 \affiliation{Departemento de F\'isica, Universidad T\'ecnica Federico Santa Mar\'ia, and Centro Cient\'ifico-\\
Tecnol\'ogico de Valpara\'iso, Avda. Espana 1680, Casilla 110-V, Valpara\'iso, Chile} 

\date{\today}


\begin{abstract}
In this paper we show, that the   $J/\psi$ production in DIS, is the main
 source of information about the events with two parton shower production.
 We attempt to develop our theoretical  acumen of this process,
 to a level compatible with the theoretical description of  inclusive
 DIS. We revisit the problem of the linear evolution equation for the double
 gluon densities, and include   Bose-Einstein enhancement to these
 equations. We find that the Bose-Einstein correlations lead to an 
increase
 of the anomalous dimension, which turns out to be suppressed as
 $1/(N^2_c -1)^2$, in  agreement with the estimates for the twist 
four
 anomalous dimension.  We believe that  understanding
 what happens to these contributions at ultra high energies, is a
 key question for an effective theory, based on high energy
 QCD. We derive the evolution equation for  the scattering amplitude of
 two dipoles with a nucleus, taking into account the shadowing
 corrections, and  investigate the analytical solutions in two
 distinct kinematic regions: deep in the saturation region, and
 in the vicinity of the saturation scale. The suggested non-linear
 evolution equation is a direct generalization of the Balitsky-Kovchegov
  equation, which has to be solved  with the initial condition that  
depends
 on the saturation scale $Q_s(Y=Y_0,b)$.
With the goal of finding a new small parameter,
 it is instructive
 to compare the solution of the non-linear equation  with
   the qusi-classical approximation, in which in the initial condition
we replace
  $Q_s(Y=Y_0,b)$ by $Q_s(Y,b)$.
Our final result is that the shadowing corrections in the 
elastic
 amplitude generate the survival probability,  which
 suppresses the  growth of the amplitude with energy,  caused by  
the Bose-Einstein enhancement.
 \end{abstract}

\preprint{TAUP - 3030/18}

\maketitle

\tableofcontents

\flushbottom

\section{Introduction}

\begin{figure}[ht]
    \centering
  \leavevmode
      \includegraphics[width=15cm]{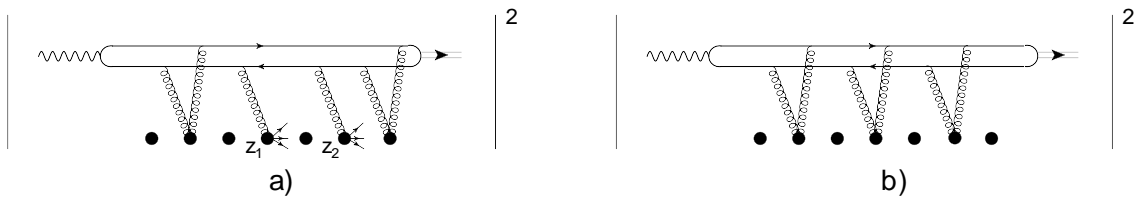}  
    \caption{ $J /\psi$ production at the initial energy
 at which the scattering amplitude is taken in the Born approximation
 of perturbative QCD.\fig{ba}-a: the  structure of the inelastic
 event; and \fig{ba}-b is  $J/\psi$  diffractive production. 
 Helical lines denote the gluons. The black blobs stand for the
 nucleons. The wavy line corresponds to the virtual photon ($\gamma^*$). }
\label{ba}
  \end{figure}

We believe that
the process for the production of the J/$\psi$ meson in the fragmentation 
region  
 in DIS,  is the simplest process  aside from inclusive DIS,
 which allows us to investigate the scattering amplitudes in the wide
 kinematic region, from    short distances  which are the subject of
 perturbative QCD, to long distances of about $1/Q_s$ ($Q_s$ is the
 saturation momentum), which can be described by the Colour Glass
 Condensate (CGC) approach(see Ref.\cite{KOLEB} for a review). For
  distances $> 1/Q_s$, the non-perturbative corrections have to
 be included. The nucleus target has  additional advantages: 
  a new parameter $\bas A^{1/3}$, which results in a  more rigorous
theoretical
approach; and   the fact that the dominant mechanism 
is  inelastic $J/\psi$ production, accompanied by the production of two
 parton showers, (see \fig{ba}), which shows the process at the  initial  
energy in the
 Born approximation of perturbative QCD, for dipole-nucleon scattering
 \cite{KLNT,DKLMT}). Therefore, we believe that this process is the main
 source of the experimental information about the event with two parton
 showers, in the same way as  inclusive DIS  is  a source for the
 structure of the one parton shower event.

The goal of this paper is  to develop our  theoretical understanding 
 of   $J/\psi$ production, to a level
 compatible with the theoretical description of  inclusive DIS. At
 high energy this  description includes the BFKL evolution
 equation\cite{BFKL}  and the CGC/saturation approach for the
 scattering amplitude\cite{GLR,MUQI,MV,MUCD,BK,JIMWLK}.  As we have 
eluded,
 the process of  $J/\psi$ production,   is intimately related
 to the double parton densities (double parton distribution functions).
 The DGLAP\cite{DGLAP} evolution equation is known for the double parton
 distribution functions (see Ref.\cite{DOS}), also the BFKL 
evolution is known
 for these processes\cite{MUPA,PESCH,LELU}. In this paper we write the 
BFKL equation
 for the particular double parton distribution that enters the $J/\psi$
 production cross sections, and we build the generating functional for all
 multi-parton distributions. 
We also
 wish   to  amend  these equations, to include the Bose-Einstein
 enhancement,
 resulting from the correlations of  identical gluons. This 
effect has  a
 suppression of the order of $1/(N^2_c-1)$, where $N_c$ is the number of 
colours, and  is closely related to the $1/(N^2_1 -1)$ corrections to the
 anomalous dimension of the twist four operator, which is larger 
 than the sum of the anomalous dimensions of two twist 2 operators, as
  has been shown in Refs.\cite{BART1,BART2,LET2}. Bose-Einstein
 correlations have drawn  considerable attention recently, since 
they give essential contributions to the azimuthal angle
 correlations\cite{BEC1,BEC2,BEC3,BEC4,BEC5,BEC6,BEC7}.  It has
 been shown (see Ref.\cite{BEC4} for example), that the Bose-Einstein
 enhancement provides a significant contribution to the measured
 angle correlations. We believe that this fact calls for a generalization
  of the evolution equation by  also including this enhancement. 
 
 Unfortunately, only in the linear approximation, is the cross section for 
 $J/\psi$ production,  determined by the double gluon density.  
 The total cross section of DIS,  is also affected by strong
 shadowing corrections\cite{GLR,MUQI,MV,MUCD,BK,JIMWLK}. We propose a
 non-linear evolution for this cross section, which is a direct 
generalization
 of the Balitsky-Kovchegov evolution. We need to solve this non-linear
 equation, using the results of Refs.\cite{KLNT,DKLMT} as the initial
 condition for this equation. It should be stressed that all formulae
 in these papers were derived using the Born approximation of perturbative
 QCD  for the scattering amplitude of a colourless dipole.
 In making estimates, the so called quasi-classical approach is widely 
used, 
replacing
 the dipole-proton cross section by $\sigma\,=\,r^2 Q^2_s/4$, where $Q_s$ 
denotes
 the saturation scale  at arbitrary $Y$, but not only at initial $Y=Y_0$,
  $r$ is the size of the dipole. This is 
certainly  not the
 correct procedure. For the total DIS cross section, this approximation 
means
 that we can use the McLerran-Venugopalan \cite{MV} formula  with the above
 replacement, instead of the correct BK non-linear equation. We can view 
our
 equation as a simplification of the general approach in the framework
 of the CGC, given in Ref.\cite{KMV},
   neglecting the $1/N_c$ correction, and using the approximations related 
to
  physical nuclei.
   
   In section VI, we show that the saturation for the elastic amplitude 
leads
 to the survival probability, which is so large that it  suppresses the 
power-like 
increase with energy, which is generated by the Bose-Einstein enhancement. 
 We summarize all our results in the Conclusions.

 
 \section{The first diagram: two BFKL Pomeron exchanges}
At present, the effective theory for QCD at high energies,   exists in two
 forms:
  the
  CGC/saturation approach \cite{GLR,MUQI,MV,MUCD,BK,JIMWLK} and the BFKL
 Pomeron
 calculus\cite{BFKL,LIREV,GLR,MUPA,BART,BRN,KOLE,LELUR,LMP} . It has been
 proven that in  general,  these two approaches are equivalent in a 
 limited region
 of the rapidities\cite{AKLL}.   However, for the Balitsky-Kovchegov 
cascade,
 which we discuss here for the interaction with the nucleus target,
 the equivalence holds in the entire kinematic region in rapidity.
  The interpretation of processes at high energy appear quite different in
 each approach,
  since  they have different  structural elements.  The CGC/saturation
 approach , being more microscopic,
  describes the high energy interactions
 in terms of colourless dipoles, their density, 
 distribution   over impact parameters, evolution in energy and so on.
 This approach can be easily applied to  inclusive processes at high
 energies, generating a new typical scale: i.e.  saturation momentum 
$Q_s$.
 
 The BFKL Pomeron calculus   which  deals with   BFKL Pomerons and
 their  
interactions, is similar to the old  Reggeon theory \cite{COL}, and is 
suitable for   describing 
 diffractive physics and correlations in multi-particle production,
 as we can use
 the Mueller diagram technique\cite{MUDI}.  The relation between different
 processes at high energy   are  very often  more transparent 
in  this approach, since in addition to the Mueller diagram technique we 
can use
 the AGK cutting rules\cite{AGK}, which are useful in spite of the 
restricted
 region of their application\cite{AGKQCD}. 
   
 In this paper we  use the BFKL Pomeron calculus for the process of  
interest, since we wish to find the relationship between the
 $J/\psi$ production and inclusive DIS. This is the main 
 difference to Ref.\cite{KMV}, in which the same process has
 been treated using the CGC approach.
 
\begin{figure}[ht]
    \centering
  \leavevmode
      \includegraphics[width=8cm]{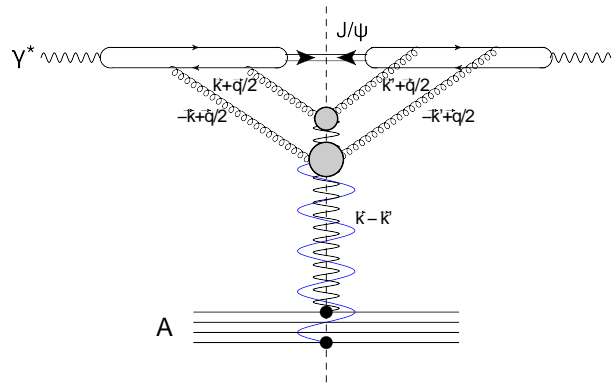}  
    \caption{The first diagram in the BFKL Pomeron calculus for 
 $J /\psi$ production. The vertical wavy lines describe the BFKL Pomerons.
 Helical lines denote the gluons. The horizontal  wavy lines correspond to 
the
 virtual photon ($\gamma^*$). The vertical dashed line shows the cut, 
which
 means that all gluons in the Pomeron are produced.}
\label{fidi}
  \end{figure}

 In the framework of the BFKL Pomeron calculus, the first diagram that
 describes  $J/\psi$ production, is the exchange of  two BFKL
 Pomerons shown in \fig{fidi}.  Since we are interested in inelastic
  $J/\psi$ production, the Pomerons in \fig{fidi} are cut Pomerons in
 which all gluons are produced.  From the unitarity constraints for the
 elastic amplitude of the dipole of size $r$, rapidity $Y$ and at
 the impact parameter $b$,
  we have
 \beq \label{UNT}
 N^{\rm BFKL}_{\rm cut}\Lb Y, r,b \Rb\,\,\equiv\,\,\sigma^{\rm BFKL}_{tot}\,
\,=\,\,2\, N^{\rm BFKL}\Lb Y, r,b \Rb 
\eeq   
 
 Its contribution to the total cross section for $J/\psi$ production
 is equal to
 \beq \label{FD1}
 \sigma\Lb Y, Q^2\Rb\,\,=\,\,\frac{\bas^2 }{(2 \pi)^6}\int d^2 k_T d^2 q_T
  d^2 Q_T \,I\Lb \vec{k}_T,\vec{q}_T\Rb\,I\Lb \vec{k}'_T,\vec{q}_T\Rb 
 N^{\rm BFKL}_{\rm cut}\Lb \vec{k}_T + \h \vec{q}_T, \vec{Q}_T\Rb \,  N^{\rm BFKL}_{\rm cut}\Lb - \vec{k}_T + \h \vec{q}_T, \vec{Q}_T\Rb 
 \eeq
 where 
  \beq \label{FD2} 
  I\Lb \vec{k}_T, \vec{q}_T\Rb\,\,=\,\, \int d^2 r \,e^{ i \h \vec{q}_T \cdot \vec{r}}\,\Big( 1\,-\,e^{ i \vec{k}_T \cdot \vec{r}}\Big)\,\, \Psi_{\gamma^*}\Lb Q , r \Rb \,\Psi_{J/\psi}\Lb r \Rb\,\,=\,\,F\Lb \vec{q}_T\Rb \,-\,F\Lb \vec{k}_T \,+\,\h \vec{q}_T\Rb
  \eeq
    and $\vec{Q}_T \,=\,\vec{k}_T \,-\,\vec{k}'_T$.  
In \eq{FD2} $\Psi_{\gamma^*}\Lb Q , r \Rb$ denotes the wave function of 
the
 dipole of size $r$ in the photon    with virtuality $Q$, while
 $\Psi_{J/\psi}\Lb r \Rb$  is the wave function of the same dipole in
 the $J/\psi$ meson.  
    The amplitude of the BFKL Pomeron $ N^{\rm BFKL}_{\rm cut}\Lb \vec{k}_T 
+ \h \vec{q}_T, \vec{Q}_T\Rb$ can be re-written in the case of the
 interaction of the Pomeron with the nucleus as follows, taking into 
account \eq{UNT}.
    \beq \label{FD3}
    N^{\rm BFKL}_{\rm cut}\Lb \vec{k}_T + \h \vec{q}_T, \vec{Q}_T\Rb   \,\,=\,\,2 N^{\rm BFKL}_N\Lb \vec{k}_T + \h \vec{q}_T, \vec{Q}_T\Rb\,S_A\Lb Q_T\Rb \,\to\,   2 N^{\rm BFKL}_N\Lb \vec{k}_T + \h \vec{q}_T, \vec{Q}_T=0 \Rb\,S_A\Lb Q_T\Rb 
    \eeq
    where $N^{\rm BFKL}_N\Lb \vec{k}_T + \h \vec{q}_T, \vec{Q}_T=0 \Rb$ 
is the scattering amplitude of the dipole with the nucleon.   
   
    \beq \label{FD4}
    S_A\Lb Q_T\Rb \,\,=\,\,\int d^2 b\,e^{i \vec{Q}_T \cdot \vec{b}}\,\int^{\infty}_{-\infty} d z \,\rho_A\Lb b,z\Rb   \,\equiv\,\int d^2 b  \,e^{i \vec{Q}_T \cdot \vec{b}}  \,S_A\Lb b\Rb
    \eeq
    where  $\rho\Lb b, z\Rb$ denotes the nucleon density in the nucleus.
 $S_A\Lb b\Rb$ is the  number  of the nucleons at fixed impact 
parameter $b$.
 The last equation in \eq{FD3} follows from the fact that the typical
 value $Q_T$ in the nucleus form factor is about $1/R_A$, where $R_A$
 denotes the radius of the nucleus. On the other hand $Q_T$ in $N^{\rm 
FKL}_N$
 is    of the order of the soft scale $\mu_{soft}$, or the saturation
 scale $Q_s$
 and, therefore, turns out to be much larger than $1/R_A$, and can be 
neglected.

    Using \eq{FD3} we can re-write \eq{FD1} in the form
     \bea \label{FD5}
 \sigma\Lb Y, Q^2\Rb\,\,&=&\,\,\frac{4 \bas^2}{(2 \pi)^4}\,\int \frac{d^2 Q_T}{(2 \pi)^2}\, S^2_A\Lb Q_T\Rb\nn\\
 &\times & 
 \int d^2 k_T d^2 q_T  \,I\Lb \vec{k}_T,\vec{q}_T\Rb\,I\Lb \vec{k}'_T,\vec{q}_T\Rb 
 N^{\rm BFKL}_{N}\Lb \vec{k}_T + \h \vec{q}_T, \vec{Q}_T=0\Rb \,  N^{\rm BFKL}_{N}\Lb - \vec{k}_T + \h \vec{q}_T, \vec{Q}_T=0\Rb\nn\\
 \,\,&=&\,\,\frac{4 \bas^2(N^2_c - 1)}{(2 \pi)^4}\,\int \frac{d^2 Q_T}{(2 \pi)^2}\, S^2_A\Lb Q_T\Rb\nn\\
 &\times & 
 \int d^2 l_T\, d^2 l'_T  \,I\Lb \vec{l}_T,\vec{l}_T + \vec{l}'_T\Rb\,I\Lb \vec{l}'_T,\vec{l}_T + \vec{l}'_T \Rb 
\, N^{\rm BFKL}_{N}\Lb \vec{l}_T , \vec{Q}_T=0\Rb \,  N^{\rm BFKL}_{N}\Lb \vec{l}'_T , \vec{Q}_T=0\Rb
 \eea    
   where $\vec{l}_T \,=\,\vec{k}_T + \h \vec{q}_T$ and $\vec{l}'_T\,=\,- 
\vec{k}_t + \h \vec{q}_T$.
   
    Introducing 
 the double transverse momentum densities which  characterize how many
 partons with 
 $(x_1,p_{1,T})$ and $(x_2,p_{2,T})$ are  in the parton cascade, and  
 which is equal to
\bea \label{FD6}
&&\rho^{(2)}\Lb x_1, \vec{p}_{1,T} ; x_2, \vec{p}_{2,T} , \vec{Q}_T\Rb\,= \,\,\sum^\infty_{n=2}\,\int\,\prod^n_{i=1}
\frac{d x_i}{ x_i}\,  d^2 k_{i,T}\,\Psi^*\Lb \{x_i,k_{i,T}\},  x_1, \vec{p}_{1,T} + \h \vec{Q}_T; x_2, \vec{p}_{2,T} - \h \vec{Q}_T \Rb\nn\\
&&\,\Bigg\{a^+(x_1, \vec{p}_{1,T} + \h \vec{Q}_T; b)\,a^+(x_2, \vec{p}_{2,T} - \h \vec{Q}_T; c)\,a(x_2, \vec{p}_{2,T} + \h \vec{Q}_T; c)\,a(x_1, \vec{p}_{1,T} - \h \vec{Q}_T; b)\Bigg\}\nn\\
&&\Psi\Lb \{x_i,k_{i,T}\},  x_1, \vec{p}_{1,T} - \h \vec{Q}_T; x_2, \vec{p}_{2,T} + \h \vec{Q}_T \Rb
\eea
where 
$\Psi\Lb \{x_i,k_{i,T}\},  x_1, \vec{p}_{1,T} - \h \vec{Q}_T; x_2,
 \vec{p}_{2,T} + \h \vec{Q}_T \Rb$ 
denotes the partonic wave function, we can re-write \eq{FD5} in the form
\beq \label{FD7}
 \sigma\Lb Y, Q^2\Rb\,\,= \,\,\frac{4 \bas^2}{(2 \pi)^2}\, \int d^2 l_T\, d^2 l'_T  \,I\Lb \vec{l}_T,\vec{l}_T + \vec{l}'_T\Rb\,I\Lb \vec{l}'_T,\vec{l}_T + \vec{l}'_T \Rb \frac{d^2 Q_T}{(2 \pi)^2} \,\,\rho^{(2)}_A\Lb x, \vec{l}_{T} ; x, \vec{l}'_{2,T} , \vec{Q}_T\Rb\nn
 \eeq
 where $Y = \ln(1/x)$.

In \eq{FD6} $a^+$ and $a$  denote the creation and annihilation operators 
for  gluons
 that have longitudinal momentum $x_i$ and transverse momentum $p_{i,T}$.
 
 For the double parton density in the coordinate representation which
 is equal to
 
 \beq \label{FD8}
 \rho^{(2)}\Lb x,\vec{r}; x, \vec{r}'; \vec{b}\Rb \,\,=\,\,\int \frac{ d^2 l_T}{(2 \pi)^2} \,\frac{ d^2 l'_T}{(2 \pi)^2} \,\frac{ d^2 Q_T}{(2 \pi)^2}\,\, e^{i \Lb\rv \cdot\vec{l}_{T}\,+\, \rv' \cdot \vec{l}'_{T}\,+\,\vec{b} \cdot \vec{Q}_T\Rb}\,
 \rho^{(2)}_A\Lb x, \vec{l}_{T} ; x, \vec{l}'_{2,T} , \vec{Q}_T\Rb 
 \eeq
 
 \eq{FD7} takes the form
 \beq \label{FD9}
 \sigma\Lb Y, Q^2\Rb\,\,= \,\,4 \bas^2 \,\, \int d^2 r\,d^2 r'\,\Phi\Lb \vec{r},\vec{r}'\Rb \int d^2 b\,  \rho^{(2)}_A\Lb x, \vec{r} ; x, \vec{r}' ;  \vec{b}_T\Rb 
 \eeq
 where
 \beq \label{FD10}
 \Phi\Lb \vec{r},\vec{r}'\Rb\,\,=\,\, \int \frac{ d^2 l_T}{(2 \pi)^2}\, \frac{d^2 l'_T }{(2 \pi)^2}\,\,e^{i \Lb \vec{l}_T \cdot \rv\,+\,\vec{l}'_T \cdot \rv'\Rb}
 \,I\Lb \vec{l}_T,\vec{l}_T + \vec{l}'_T\Rb\,I\Lb \vec{l}'_T,\vec{l}_T + \vec{l}'_T \Rb 
  \eeq

   
 \section{Generating functional  and  BFKL evolution for the
 multi-gluon densities}
 
 The evolution equations for the multi-parton densities has been
 given in Ref.\cite{LELU}. In this section we briefly review this 
proof, and derive the non-linear evolution equation for the generating 
functional with two colourless dipoles   as the initial condition.   This
 will give us  the linear BFKL evolution  for all partonic densities
 $\rho^{(n)}\Lb\{\vec{r}_i,\vec{b}_i\}\Rb$. Following Ref.\cite{MUCD} 
we consider a colourless dipole of size $\vec{r}_i$  at  impact
 parameter $\vec{b}_i$, as the parton in the fast projectile which can be
 described by the partonic wave function $\Psi\Lb \{\vec{r}_i,\vec{b}_i\}\Rb$. 
 For a system of $n$-partons with coordinates $\{ \vec{r}_i, \vec{b}_i\}$
 with $i = 1,\dots ,n$ and rapidity $Y$ ,  we can introduce the probability
 (probability density)  to find  them in the projectile wave function:
 $P_n\Lb Y;\{ \vec{r}_i, \vec{b}_i\} \Rb$.  For $P_n$ we write the
 classical cascade equation\cite{MUCD,LELU,LELUR}:
 
 \begin{eqnarray}\label{PDE1}
&&\,\,\frac{\partial\,P_n\left(Y\,-\,Y_0;\{\vec{r}_i, \vec{b}_i\} \right)}{ 
\bar{\alpha}_s\,\partial\, Y}\,=\nn\\
&&\,-\,
\sum^n_{i=1}\,\omega(r_i) \,
P_n\left(Y\,-\,Y_0; \{\vec{r}_i, \vec{b}_i\} \right)
+\,\sum^{n-1}_{i=1} \,\frac{(r_i\,+\,r_n)^2}{(2\,\pi)\,r^2_i\,r^2_n}\,
P_{n - 1}\left(Y\,-\,Y_0; \{\vec{r}_i, \vec{b}_i\}\;  \vec{r}_i \to \vec{r}_i + \vec{r}_n, \vec{b}_i \to \vec{b}_{in}
 \right)
\end{eqnarray} 
 with $\vec{b}_{in}\,=\,\vec{b}_i\,+\,\vec{r}_n/2\,=\,\vec{b}_n\,-\,\vec{r}_i/2$.
The two terms of Eq.(\ref{PDE1}) have a 
 simple meaning: the first one describes
the decrease in probability to find $n$ dipoles, due to a decay of one 
dipole
into two dipoles  of arbitrary size.  This probability is equal to
\beq \label{PDE2}
 \bar{\alpha}_s \,\,\omega(r_i)\,\,=\,\,\frac{ 
\bas}{2\,\pi}\,\int_\rho 
\,\frac{r_i^2}{(r_i\,-\,r')^2\,r'^2}\,d^2 r'\,\,=\,\,
\bas \,\,\ln(r_i^2/\rho^2)
\eeq
where  $\rho$ is an  infrared cutoff.
The second term, reflects  the increase in probability to find $n$  
dipoles
due to a creation of a new dipole from  $n -1$ dipoles, with probability
\beq \label{PDE3}
 \frac{\bas}{2\,\pi} \,\,\frac{(r_1 \,\,+\,\, 
r_2)^2}{r_1^{2}\,r_2^2}\,\,.
\eeq
To find the probabilities,
 we need  to add  the initial condition at $Y = Y_0$ to  \eq{PDE1}. 
  We wish to stress that we are only dealing
  with the decay of the dipoles, neglecting the possibilities for the
 dipoles to recombine. In terms of the Pomeron calculus, it means that
 we only take into account  `fan' Pomeron diagrams, which give the
 dominant contribution for  DIS with the nuclear target\cite{GLR,MUCD,BK}.
 
 \eq{PDE1} can be re-written in more elegant form 
introducing a generating functional $Z$
\beq \label{PDE4}
Z\left(Y\,-\,Y_0;\,[u_i] \right)\,\,\equiv\,\,\sum_{n=1}\,\int\,\,
P_n\left(Y\,-\,Y_0;\,\{\vec{r}_i,\vec{b}_i\}
 \right) \,\,
\prod^{n}_{i=1}\,u(r_i, b_i) \,d^2\,r_i\,d^2\,b_i
\eeq 
where $u(r_i, b_i) \equiv u_i $ is an arbitrary function of $r_i$ and $b_i$. 
It  follows immediately from \eq{PDE4}
that the functional obeys the condition:
at $u_i\,=\,1$ 
\beq \label{LDIN2} 
Z\left(Y\,-\,Y_0;\,[u_i=1]\right)\,\,=\,\,1\,.
\eeq
The physical meaning of (\ref{LDIN2}) is that the sum over
all probabilities is one.

Multiplying \eq{PDE1} by the product $\prod^n_{i=1}\,u_i$ 
and integrating over all $r_i$ and $b_i$,  we obtain the 
following linear equation for the generating functional:
\beq \label{LEQZ}
\,\frac{\partial \,Z}{\bas\,\partial \,Y}\,\,=\,\,-\,\,
\int\,d^2 r\,d^2 b\,\,V_{1\rightarrow 1}(\rv ,\,\bv,\,[u])\,\, Z\,\,
+\,\,\int\,\,d^2 \,r \,d^2\,r' \,d^2 b\,\,V_{1\rightarrow 2}(\rv ,\,\rv', \,\bv,\,[u])\,\, Z\,
\eeq
with the definitions
\beq \label{V2} 
V_{1 \rightarrow 1}(\rv ,\, \bv,\,[u])\,\,=\,\,
\bas \,\,\omega(r) \,\,u(\vec{r},\,\vec{b})\,\,\frac{\delta}{\delta u(\vec{r},\vec{b})}
\eeq
and
\beq \label{V1}
V_{1 \rightarrow 2}(\rv,\,\rv',\,\bv,\,[u])\,\,
=\,\,\frac{ \bas}{2\,\pi} 
\,\,\frac{r^2}{r'^2\,(\vec{r} -\vec{ r}')^2}\,\,
\,u\Lb\vec{r}', \,\vec{b}\,+\,\h(\rv\,-\,\rv')\Rb\,\,u\Lb\rv \,-\,\rv', \,\bv\,-\,\h \rv\Rb\,\,
\frac{\delta}{\delta u(\rv, \,\bv)}\,.
\eeq 
The functional derivative with respect to $u(r,b)$,  plays the  role 
of an  annihilation operator for a dipole of  size $r$,  at  impact 
parameter $b$. 
The multiplication by $u(r,b)$ corresponds to
a creation operator for this dipole. Having this in mind, we can see  that
the $n$-dipole densities in the projectile 
$\rho^{(n)}\Lb Y - Y_0; \{ \rv_i,\bv_i\}\Rb$
 are defined as follows:
\beq \label{RON}
\rho^{(n)}\Lb Y - Y_0; \{ \rv_i,\bv_i\}\Rb\,=\,\frac{1}{n!}\,\prod^n_{i =1}
\,\frac{\delta}{\delta
u_i } \,Z\left(Y\,-\,Y_0;\,[u_i] \right)|_{u=1}
\eeq
Using \eq{RON} we  obtain the following linear evolution equations
 for  $\rho^{(n)}\Lb Y - Y_0; \{ \rv_i,\bv_i\}\Rb $:
\bea\label{EVRON}
&&\frac{\partial \,\rho^{(n)}(Y - Y_0; \rv_1, \bv_1\,\ldots\,,\rv_n, \bv_n)}{ 
\bas\,\partial\,Y}\,\,=\,\,
-\,\sum_{i=1}^n
 \,\,\omega(r_i)\,\,\rho^{(n)}( Y - Y_0; \rv_1, \bv_1\,\ldots\,,\rv_n, \bv_n)\,\,\nn\\
 &&+
2\,\sum_{i=1}^n\,
\int\,\frac{d^2\,r'}{2\,\pi}\,
\frac{r'^2}{r^2_i\,(\rv_i\,-\,\rv')^2}\,
\rho^{(n)}(\ldots\,\rv', \bv_i- \h\rv', \dots)\, 
+\,\sum_{i=1}^{n-1}\,\frac{(\rv_i + \rv_n)^2}
{(2\,\pi)\,r^2_i\,r^2_n}\,
\rho^{(n-1)}(\ldots\,(\rv_i\,+\,\rv_n), \bv_{in},\dots)
 \eea
For $ r^2_1\,r^2_2\,\int d^2 b\,  \rho^{(2)}_A\Lb x, \vec{r} ; x, \vec{r}' ; 
 \vec{b}_T\Rb \,=\,\tilde{\rho}^{(2)}\Lb Y - Y_0, \rv_1,\rv_2\Rb$ \eq{EVRON}
 can be re-written in the following form

\bea \label{EVRONTI}
&&\frac{\partial \,\tilde{\rho}^{(2)}(Y - Y_0; \rv_1, \rv_2)}{ 
\bas\,\partial\,Y}\,=\\
&&\,\sum_{i=1}^2\int \frac{d^2 r'}{2 \pi} \frac{1}{\Lb \rv_i - \rv'\Rb^2} \Bigg\{ 2 \tilde{\rho}^{(2)}(Y - Y_0; \rv', \rv_{i+1})
 \,-\,\frac{r^2_1}{r'^2}\tilde{\rho}^{(2)}(Y - Y_0; \rv_1, \rv_2) \Bigg\}\,\,+\,\,\tilde{\rho}^{(1)}\Lb  Y - Y_0,  \rv_1+\rv_2\Rb\nn
 \eea 
where $ \Lb \rv_1 + \rv_2\Rb^2\int d^2 b \,\rho^{(1)}\Lb  Y - Y_0,
  \rv_1+\rv_2, \bv \Rb \,\,=\,\, \tilde{\rho}^{(1)}\Lb  Y - Y_0,  
\rv_1+\rv_2\Rb$.

In the  momentum representation \eq{EVRONTI} takes the  form which is similar
 to the DGLAP evolution ( see Ref.\cite{DOS} and the references  therein):
\beq \label{EVRONMO}
\frac{\partial \,\tilde{\rho}^{(2)}(Y - Y_0; \pv_{1,T}, \pv_{2,T})}{ \bas
\,\partial\,Y}\,\,\,=\,\,\,\sum_{i=1}^2\int \frac{d^2  k_T }{\pi}  \,K\Lb \pv_i , \kv\Rb  \,\, \tilde{\rho}^{(2)}(Y - Y_0; \kv_{i,T}, \pv_{i+1})
\,\,+\,\,\tilde{\rho}^{(1)}\Lb  Y - Y_0,  \pv_{1,T}\Rb\delta^{(2)}\Lb \vec{p}_{1,T} - \vec{p}_{t,2}\Rb\nn
 \eeq 
where $K\Lb \pv_i , \kv\Rb $ denotes the BFKL kernel that has the form

\bea \label{KMO}
&&K\Lb \pv_i , \kv\Rb\,\tilde{\rho}^{(2)}(Y - Y_0; \kv_{i,T}, \pv_{i+1,T})\,\,=\nn\\
&&\,\,\frac{1}{\Lb \pv_{i,T} - \kv_T\Rb^2}\Bigg\{  \tilde{\rho}^{(2)}(Y - Y_0; \kv_{i,T}, \pv_{i+1,T})\,\,-\,\,\frac{p^2_{i,T}}{k^2_T\,+\,\Lb \pv_{i,T} - \kv_T\Rb^2}  \tilde{\rho}^{(2)}(Y - Y_0; \pv_{i,T}, \pv_{i+1,T})\Bigg\}\eea

\eq{EVRON} can be resolved  if we know the initial conditions. 
Assuming that at $Y = Y_0$, $\rho^{(1)} = \delta^{(2)}\Lb \rv - \rv_1\Rb\,
\delta^{(2)}\Lb \bv - \bv_1\Rb$ while $\rho^{(n)} = 0$ for $n>1$,
 we see that at $Y = Y_0$ the initial condition for the generating 
functional implies
\beq \label{INCON1}
Z^{(1)}\Lb Y = Y_0, \{\rv_i,\bv_i\},[u_i]\Rb \,=\,u\Lb \vec{r}, \vec{b}\Rb
\eeq
The linear equation  \eq{LEQZ} has a general solution of the form
 $Z\Lb Y - Y_0, \{\rv_i,\bv_i\},[ u_i]\Rb \,=\,Z\Lb \{u_i\Lb Y-Y_0,\rv_i,\bv_i\Rb\}\Rb$ and using \eq{INCON1} we obtain the non-linear equation of Ref.\cite{MUCD}:
\bea \label{NLEQZF}
&&\frac{\partial\, Z^{(1)}\left(Y, \,r,\, b;\,[u] \right) }{
\bas\,\partial \,Y}\,\,=\,\,- \,\,\omega(r)\,\,
Z^{(1)}\left(Y; \,r,\, b;\,[u] \right)\nn\\
&&
+\,\,\int\,\,
\frac{d^2\,r'}{2\,\pi}\,\,\frac{r^2}{r'^2\,(r\,-\,r')^2}\,\,
Z^{(1)}\left(Y;\,r',\, b\,+\,\frac{(r\,-\,r')}{2};\,[u] \right)\,\,
Z^{(1)}\left(Y;\,(r\,-\,r'),\, b\,-\,\frac{r'}{2};\,[u] \right)\,.
\eea

As we have demonstrated for the $J/\psi$ production, the   
initial
 conditions at $Y = Y_0$ are  $\rho^{(2)} = \delta^{(2)}\Lb \rv - 
\rv_1\Rb\,\delta^{(2)}\Lb \bv - \bv_1\Rb\,\, \delta^{(2)}\Lb \rv' -
 \rv_2\Rb\,\delta^{(2)}\Lb \bv' - \bv_2\Rb$\,\,and all $\rho^{(n)} = 0
 $ for $n \neq 2 $. In other words, for $Z$ we have the initial condition

\beq \label{INCO21}
Z^{(2)}\Lb Y = Y_0, \{\rv_i,\bv_i\},[u_i]\Rb \,=\,u\Lb \vec{r}_1,
 \vec{b}_1\Rb\,u\Lb \vec{r}_2, \vec{b}_2\Rb\eeq

Using this equation we obtain

\bea \label{NLEQZF2}
&&\frac{\partial\, Z^{(2)}\left(Y, \,\rv_1,\, \bv_1; \rv_2,\bv_2;\,[u]
 \right) }{
\bas\,\partial \,Y}\,\,=\,\,- \,\,\Lb \omega(r_1)\,\,+\,\,\omega(r_2)\Rb
Z^{(2)}\left(Y, \,\rv_1,\, \bv_1; \rv_2,\bv_2;\,[u] \right)\nn\\
&&+
\,\,\int\,\,
\frac{d^2\,r'}{2\,\pi}\,\,\frac{r_1^2}{r'^2\,(\rv_1\,-\,\rv')^2}\,\,
Z^{(1)}\left(Y;\,r',\, b\,+\,\h(\rv_1\,-\,\rv');\,[u] \right)\,\,
Z^{(2)}\left(Y;\,\rv_1\,-\,\rv',\, \bv_1\,-\,\h \rv'; \rv_2, \bv_2; \,[u]\right)\,\,\nn\\
&&+\,\,\int\,\,\frac{d^2\,r'}{2\,\pi}\,\,\frac{r_1^2}{r'^2\,(\rv_2\,-\,\rv')^2}\,\,
Z^{(1)}\left(Y;\,r',\, b\,+\,\h(\rv_2\,-\,\rv');\,[u] \right)\,\,
Z^{(2)}\left(Y;\,\rv_2\,-\,\rv',\, \bv_2\,-\,\h \rv'; \rv_1, \bv_1;[u]\right)
\eea

\section{Bose-Einstein enhancement in the evolution of  double parton
 densities }

\begin{figure}[ht]
    \centering
  \leavevmode
      \includegraphics[width=13cm]{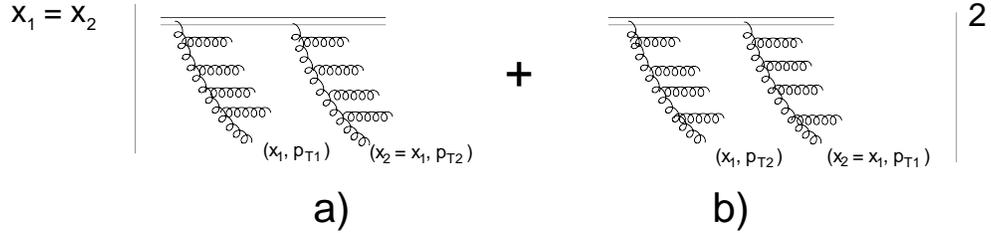}  
    \caption{ The origin of  interference diagrams  which violate
 \eq{BE1}, and should  be included in the evolution. }
\label{be}
  \end{figure}

 Comparing \eq{FD6},\eq{FD7} and \eq{FD8}, one can see that the double
 gluon density has a general form given by
 \beq \label{BE1}
\rho^{(2)}\Lb x_1, \vec{p}_{1,T} + \h \vec{Q}_T; x_2, \vec{p}_{2,T} - \h \vec{Q}_T\Rb\,=\,F\Lb Q_T\Rb \,\rho^{(1)}\Lb x_1,p_{1,T}\Rb\,\rho^{(1)}\Lb x_2,p_{2,T}\Rb
\eeq 

\eq{BE1}  indicates that two gluons are produced from two different 
parton showers,
   and have no correlations between them, at least at large values of 
rapidities.
 However, we  can see from \fig{be}  that
 \eq{BE1} cannot be correct, due to the interference between 
gluons coming from different single parton densities, which is possible, 
and  could
 give sufficiently large contributions for the identical  
gluons\cite{BEC4}.
 
  \subsection{The interference diagram in the DLA}
 In this section  for a sake of simplicity,  we will treat the
 interference diagrams in double log approximation(DLA) of perturbative
 QCD, in which we consider $\bas Y \geq 1, \bas \ln\Lb p^2_T\Rb  \geq 1$,
 but $\bas \ll 1$. This approximation appears  more  credible not 
for 
the
 gluon densities, but for double gluon distribution function which is 
defined
 as
 \beq \label{ID1}
 D^{(2)}\Lb x_1, x_2, Q\Rb \,=\, \int^Q \frac{d^2 p_{1,T}}{(2 \pi)^2} \,\int^Q  \frac{d^2 p_{2,T}}{(2 \pi)^2}  \int^Q  \frac{d^2 Q_{T}}{(2 \pi)^2}  \,\, \rho^{(2)}\Lb x_1, \vec{p}_{1,T} + \h \vec{Q}_T; x_2, \vec{p}_{2,T} - \h \vec{Q}_T\Rb 
 \eeq
 At large $Q$, the cross section of  $J/\psi$ production depends
 on $D^{(2)}\Lb x, x, Q\Rb$, as can be seen from \eq{FD8}.
  
\begin{figure}[ht]
    \centering
  \leavevmode
  \begin{tabular}{l l}
      \includegraphics[width=13cm]{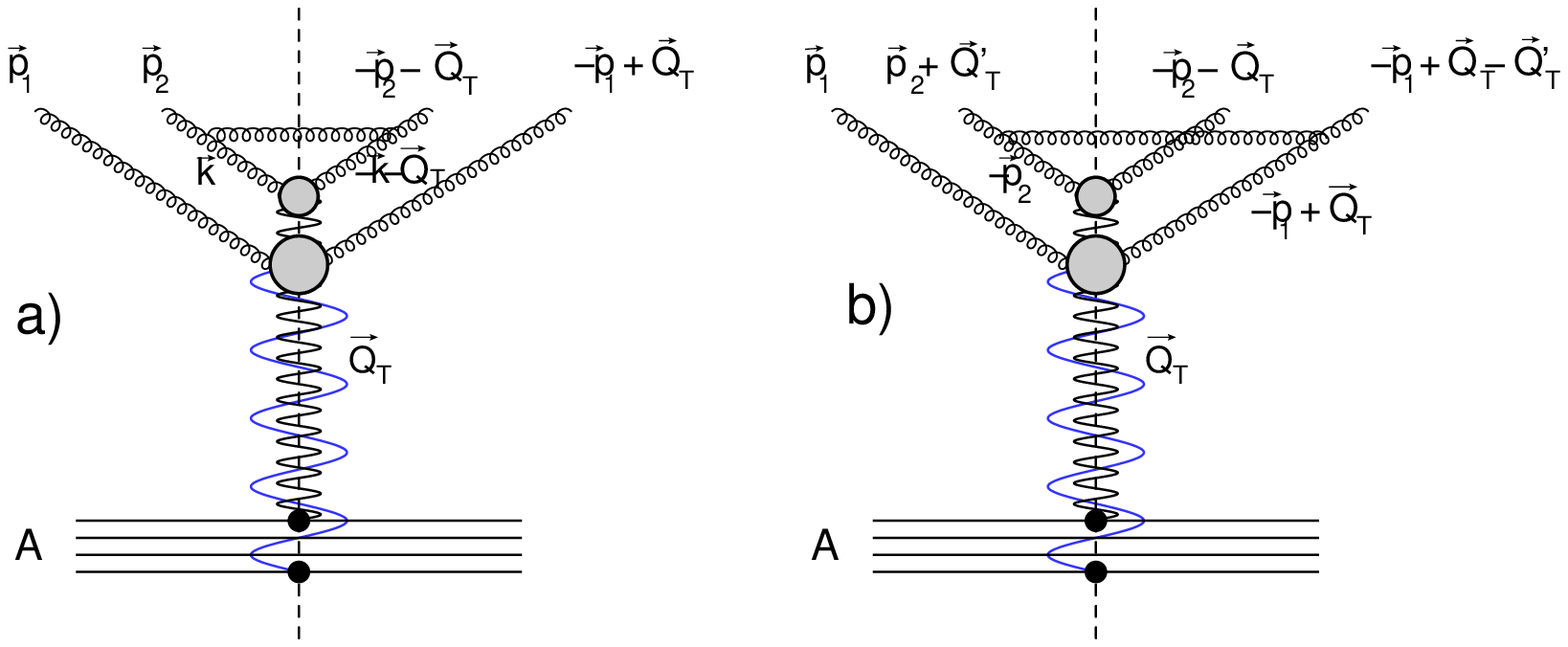} & \includegraphics
[width=5cm]{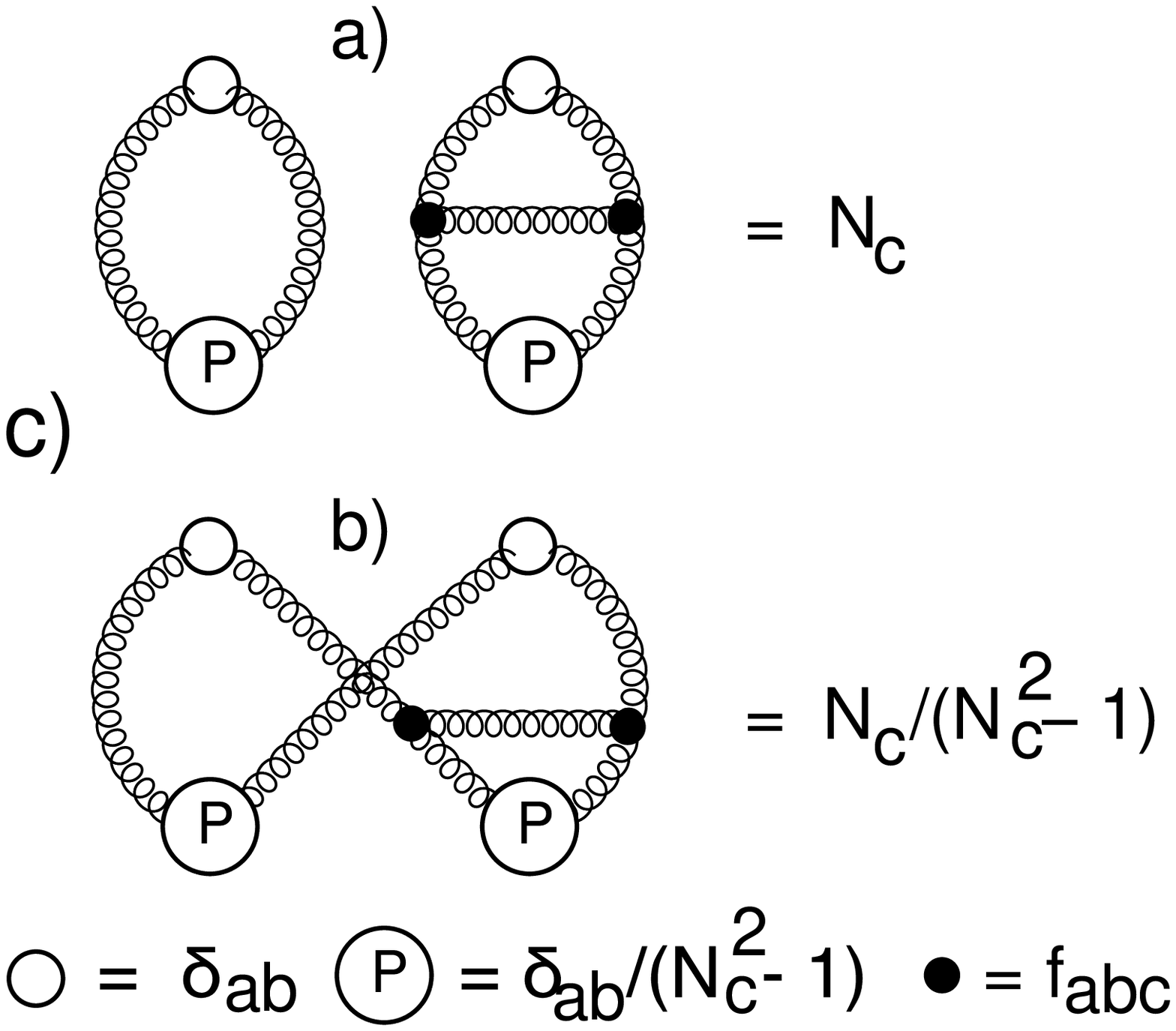}\\
      \end{tabular}    \caption{ The first interference diagram for
 double gluon densities (\fig{indi}-b) in DLA of perturbative QCD.
 \fig{indi}-a shows the emission an extra gluon in one of the parton
 showers. \fig{indi}-c shows the calculation of the colour coefficient
 in the diagrams: \fig{indi}-c a for \fig{indi}-a and \fig{indi}-c b for
 \fig{indi}-b.}
\label{indi}
  \end{figure}

First, we consider the diagram of \fig{indi}-a which shows the emission
 of an extra gluon in one parton shower. This emission leads to the DLA
 increase of $ \rho^{(1)}\Lb x_2,\vec{p}_{2,T}\Rb =
 \rho^{(1)}\Lb Y_2,\vec{p}_{2,T}\Rb$. 
Recalling that  $Q_T \propto 1/R_A \,\ll \,p_{i,T}$  and
   ( see Ref.\cite{KOLEB} for example)
\bea  
1.~~~~~\rho^{(1)}\Lb Y_2=0; \vec{p}_{2,T}; \vec{p}_{2,T} - \vec{Q}_T \Rb\,\,&=&\,\,\bas\,\delta_{a,b}\frac{\vec{p}_{2,T} \cdot\vec{p}_{2,T} + \vec{Q}_T}{p^2_{2,T}\,\Lb \vec{p}_{2,T} + \vec{Q}_T\Rb^2}\,S_A\Lb Q_T\Rb;\label{ID21}\\
2.~~~~~\Gamma_\mu\Lb \vec{p}_{2,T},\vec{k}_T\Rb &\xrightarrow{p_{2,T}  \gg k_T}&  2 g \,f^{abc} \vec{p}_{2,T};\label{ID22}
\eea
 we see that 
 \beq \label{ID3}
  \rho^{(1)}\Lb Y, \vec{p}_{2,T} ; Y, \vec{p}_{2,T} + \vec{Q}_T\Rb\,=\, 
 \rho^{(1)}\Lb Y, \vec{p}_{2,T} ; Y, \vec{p}_{2,T} \Rb\,=\,\int^Y d Y'\,
\frac{\bas}{p^2_{2,T}} \int^{p^2_{2,T}} d k^2_T \, \rho^{(1)}\Lb Y', 
\vec{k}_{T} ; Y,' \vec{k}_{T} \Rb   
  \eeq

 For $D^{(1)}\Lb Y, Q\Rb $ \eq{ID3} reduces to
 \beq \label{ID4}
 D^{(1)}\Lb Y, Q\Rb\,\,=\,\, \bas\int^Y d Y'\, \int^Q \frac{d^2 p_{2,T} }{p^2_{2,T} } \, D^{(1)}\Lb Y', p_{2,T}\Rb
 \eeq
 
 For $D^{(2)}$ the sum of the diagrams of \fig{indi}-a give
 \beq  \label{ID5}
  \frac{ \partial D^{(2)}\Lb Y, Y, \xi; {\rm extra\,\, gluon}\Rb}{\partial\,
 \xi}\,\,=\,\, \,\Bigg\{\int^Y d Y'\, D^{(2)}\Lb Y', Y,\xi\Rb 
  + \int^Y d Y'\, D^{(2)}\Lb Y, Y',\xi\Rb  \Bigg\} 
  \eeq
  where $\xi = \bas \ln\Lb Q^2\Rb$.
  
  We wish to  stress that $D^{(2)}\Lb Y, Y, \xi; {\rm extra\,
\, gluon}\Rb$ can be calculated only using 
 $ D^{(2)}\Lb Y,' Y, \xi\Rb  $ at different values of $Y_1 (x_1)$ and 
$Y_2(x_2)$.
    
  The interference diagram of \fig{indi}-b can be written in the following form:
  
  \bea \label{ID6}
&&  D^{(2)}_{\rm int}\Lb Y, Y, \xi; {\rm extra\,\, gluon}\Rb\,\,=\,\nn\\
  &&\,\frac{\bas}{N^2_c - 1} \int^Y d Y'  \int \frac{d^2 p_{1,T}}{(2 \pi)^2} \,\int  \frac{d^2 p_{2,T}}{(2 \pi)^2}  \int \frac{d^2 Q'_{T}}{(2 \pi)^2}\int \frac{d^2 Q_{T}}{(2 \pi)^2} \,\frac{\Lb \vec{p}_{2,T} + \vec{Q}'_T \Rb\cdot\Lb 
  \vec{p}_{1,T} - \vec{Q}_T + \vec{Q'}_T\Rb}{\Lb  \vec{p}_{2,T} + \vec{Q}'_T \Rb^2\,\,\Lb 
  \vec{p}_{1,T} - \vec{Q}_T + \vec{Q'}_T\Rb^2}\,\,  \nn\\
 &&\times\,\, \rho^{(1)}\Lb Y', \vec{p}_{1,T} ; Y,' \vec{p}_{1,T} - \vec{Q}_T\Rb   \rho^{(1)}\Lb Y', \vec{p}_{2,T} + \vec{Q}_T ; Y,' \vec{p}_{2,T} \Rb    
  \eea 
  where the colour coefficient reflects the fact that only identical 
gluons
  contribute to this diagram. The calculation of the colour
 coefficients are shown in \fig{indi}-c.
  
  One can see that for $Q_T\, \ll \,p_{1,T} \,\leq\,Q'_T\, \leq\, Q$
 and for $Q_T\,\ll\,p_{2,T} \,\leq\,Q'_T\,\leq\,Q$
  \eq{ID6} can be rewritten in the form
  \beq \label{ID7}
   D^{(2)}_{\rm int}\Lb Y, Y, \xi; {\rm extra\,\, gluon}\Rb
  \,\,=\,\,\bas \int^Y d Y'\,\int^Q\frac{d Q'^2_T}{Q'^2_T}
   D^{(2)}_{\rm \fig{fidi}}\Lb Y, Y, Q'_T \Rb 
   \eeq 
  We would like to draw the reader's attention to the integration over 
$Q_T$ and
 $Q'_T$, since,  as has been noticed in Refs.\cite{BART1,BART2,LET2},
 without this integration the interference diagrams will not acquire 
 the term proportional to $\ln(Q^2)$\footnote{We thank Jochen Bartels
 for discussion on the importance of $Q_T$ integration.}.

  \subsection{DGLAP evolution with  Bose-Einstein enhancement:
 anomalous dimension}
   
    The evolution equation for the  double parton distribution function 
 can be derived from \eq{EVRONMO},and in the DLA
    and it has the following  form:
    \beq \label{EQBE1}
\frac{\partial D^{(2)}\Lb Y, Y,Q\Rb}{\partial \,\xi} \,\,=\,\, \int^{Y}_{Y_0} d Y' \,\,D^{(2)}\Lb Y' ,Y,Q\Rb\,\,+\, \int^{Y}_{Y_0} d Y'\,\,D^{(2)}\Lb Y, Y', Q\Rb\,\,+\,\,\,D^{(1)}\Lb Y, Q\Rb  
 \eeq  
  Note that, the last term in \eq{EQBE1}  depends on
 $x_1+x_2 = 2 x$, but in the leading  $log (1/x) $ approximation, we do 
not  
take into account the difference between $ \tilde{Y} = \ln \Lb1/(2x)\Rb
 $ and $Y = \ln(1/x)$, on which  the first two terms depend.
 
\begin{figure}[ht]
    \centering
  \leavevmode
      \includegraphics[width=16cm]{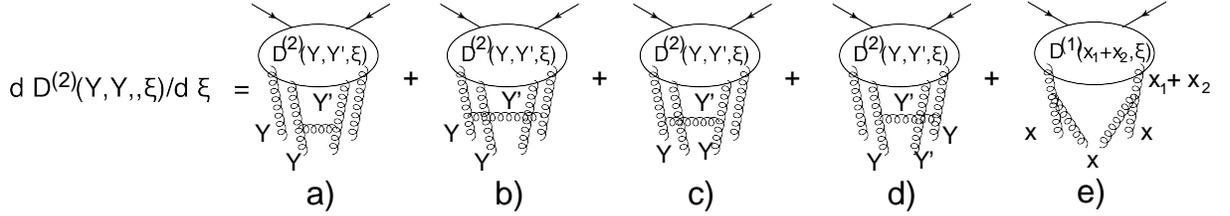}  
    \caption{ DGLAP evolution equation for the double parton distribution
 with the Bose-Einstein enhancement.}
\label{dglapf}
  \end{figure}

     To include the Bose-Einstein enhancement, we 
need  a more general evolution equation, which is shown in
 \fig{dglapf}. In this equation we add to \eq{EQBE1} the term,
 which is given by     
      \eq{ID7},  and which   describes the contributions of the diagrams
 of \fig{dglapf}-c and \fig{dglapf}-d. In the DLA this  equation has 
the form: 
\beq \label{EQBE2}
\frac{\partial D^{(2)}\Lb Y, Y,\xi\Rb}{\partial \,\xi} \,\,=\,\, \int^{Y}_{Y_0}\!\!\!\! d Y' \,\,D^{(2)}\Lb Y' ,Y,\xi\Rb\,\,+\, 
\int^{Y}_{Y_0} \!\!\!\! d Y'\,\,D^{(2)}\Lb Y, Y', \xi\Rb\,\,+\,\,\frac{2}{N^2_c - 1}\int^Y\!\!\!\!  d Y'\,D^{(2)}\Lb Y', Y',\xi\Rb\,\,+\,\,D^{(1)}\Lb Y, \xi\Rb  
 \eeq      
 Calculating the colour coefficient (see \fig{indi}-c), we assumed that
 the main
 contribution in the region of small $x$, comes from the colourless
 states in the $t$-channel for the gluons with the same values of
 $x_i$, as  is shown in \fig{indi}. In \eq{EQBE2} we use a new
 notation $Y \,=\,2 \,N_c \ln(1/x)$.

   We  first solve the    homogeneous equation:
 
   \beq \label{EQBE3}
\frac{\partial D^{(2)}\Lb Y, Y,\xi\Rb}{\partial \,\xi} \,\,=\,\, \int^{Y}_{Y_0}\!\!\!\! d Y' \,\,D^{(2)}\Lb Y' ,Y,\xi\Rb\,\,+\, 
\int^{Y}_{Y_0} \!\!\!\! d Y'\,\,D^{(2)}\Lb Y, Y', \xi\Rb\,\,+\,\,\frac{2}{N^2_c - 1}\int^Y\!\!\!\!  d Y'\,D^{(2)}\Lb Y', Y',\xi\Rb 
 \eeq    
 
   We introduce the following Mellin transform to solve this equation:
 \bea \label{EQBE4}
 D^{(2)}\Lb Y_1, Y_2,\xi\Rb\,\,&=&\,\,\int^{\epsilon + i \infty}_{\epsilon - i \infty}\frac{d \gamma}{2 \pi i}
\int^{\epsilon + i \infty}_{\epsilon - i \infty}\frac{d \omega_1}{2 \pi i}\int^{\epsilon + i \infty}_{\epsilon - i \infty}\frac{d \omega_2}{2 \pi i}\,\,e^{  \gamma \xi + \omega_1 Y_1 + \omega_2 Y_2}\,d^{(2)}\Lb \omega_1,\omega_2,\gamma\Rb\\
 &=& 2\, \int^{\epsilon + i \infty}_{\epsilon - i \infty}\frac{d \gamma}{2 \pi i}
\int^{\epsilon + i \infty}_{\epsilon - i \infty}\frac{d \omega_\Sigma}{2 \pi i}\int^{\epsilon + i \infty}_{\epsilon - i \infty}\frac{d \omega_D}{2 \pi i}\,\,e^{  \gamma \xi + \omega_\Sigma \Lb Y_1 + Y_2\Rb + \omega_D\Lb Y_1 - Y_2\Rb}\,d^{(2)}\Lb \omega_\Sigma,\omega_D,\gamma\Rb\nn
\eea 
where $\omega_\Sigma = \h (\omega_1 + \omega_2)$
 and $\omega_D = \h (\omega_1 - \omega_2)$.

 Plugging \eq{EQBE4} into \eq{EQBE3} we obtain
 \beq \label{EQBE5}
 \gamma d^{(2)}\Lb \omega_\Sigma,\omega_D,\gamma\Rb\,=\,\Bigg(\frac{1}{\omega_\Sigma + \omega_D}\,+\,\frac{1}{\omega_\Sigma - \omega_D}\Bigg) \,+\,\delta \int^{\epsilon + i \infty}_{\epsilon - i \infty}\frac{d \omega_D}{2 \pi i}  \, d^{(2)}\Lb \omega_\Sigma,\omega_D,\gamma\Rb \eeq
 where $\delta = 2/(N^2_c - 1)$.

   Looking for a solution  of the form
   \beq \label{SOLBE}
   d^{(2)}\Lb \omega_\Sigma,\omega_D,\gamma; q_T\Rb\,\,=\,\,\frac{ d^{(2)}_{in}\Lb \omega_\Sigma; q_T\Rb}{ \Lb\gamma\Lb \omega^2_\Sigma\,-\,\omega^2_D\Rb \,-\,2\omega_\Sigma\Rb}
   \eeq
   we obtain  for the spectrum of the linear equation:
   \beq \label{SOLBE1}
1\,\,=\,\,\h\delta\frac{1 }{\sqrt{\gamma\,\omega_\Sigma \,\,\Lb \gamma \omega_\Sigma\,-\,2\Rb}}
\eeq
 Utilizing the smallness of $\delta$, we see that  we have an
anomalous dimension, whose value is larger than at  $\delta =0$. 
This fact has been noted and 
discussed,
 for the twist four anomalous dimension, in 
Refs.\cite{BART1,BART2,LET2,BARY,LLS,LALE,BARY1}. The solution to
 \eq{SOLBE1} at $\omega_D  = 0$  has the following form:
\beq \label{SOLBE2}
\gamma_{an}\Lb \omega_\Sigma, \omega_D=0\Rb\,\, = \,\,\frac{2}{\omega_\Sigma}\Bigg( 1 \,+\,\frac{\delta^2}{16}\Bigg)
\eeq
 One  can 
 check that the general form of the anomalous dimension  is
\beq \label{SOLBE3}
\gamma_{an}\Lb \omega_1, \omega_2\Rb\,\, = \,\frac{1}{\omega_1}\,\,+\,\,\frac{1}{\omega_2}\,\,+\,\, \frac{\delta^2}{8}\,\,\frac{2}{\omega_1\,+\,\omega_2} \,\,\equiv\,\,\,\frac{1}{\omega_1}\,\,+\,\,\frac{1}{\omega_2}\,\,+\,\,\tilde{\delta}\,\frac{1}{\omega_1\,+\,\omega_2}
\eeq   
with $\tilde{\delta}= 1\Big{/}\Lb N^2_c - 1\Rb^2$.

We wish to  stress, that the corrections to the anomalous
 dimension of the double parton distributions, are of the order
 $\Lb \frac{1}{N^2_c - 1}\Rb^{2}$. It has the same dependence on $N_c$,
 as the  correction to the anomalous dimension of 
the twist four
 operator \cite{BART1,BART2,LET2}.

 The numerical value for $\tilde{\delta}$
  in \eq{SOLBE3} is about 0.016 for $N_c=3$, which indicates   that 
this  correction is rather small.  However, these corrections generate
 the term which increases both with $Y$ and $\xi$, and could be important
 in the region of small $x$. We will examine them in the next section.

 The  procedure just described, is a method to find the first 
corrections
 with respect to $\delta$,  to the value of the anomalous dimension.
 Replacing  \eq{SOLBE} by the following expression
\beq \label{SOLBE4}
   d^{(2)}\Lb \omega_\Sigma,\omega_D,\gamma\Rb\,\,=\,\,\frac{ d^{(2)}_{in}\Lb \omega_
\Sigma\Rb}{ \Lb\gamma \,-\,\gamma^{(1)}_{an}\Lb \omega_1,\omega_2\Rb\Rb\,
\Lb \omega^2_\Sigma\,-\,\omega^2_D\Rb}
   \eeq
 we obtain an equation for the next order correction to the value of the
 anomalous dimension. It has the form
 \beq \label{SOLBE5}
 1\,\,=\,\,\h\delta\frac{\gamma\,\omega_\Sigma \,- 
\,1 }{\sqrt{\gamma\,\omega_\Sigma}\ \,\,\Lb \gamma 
\omega_\Sigma\,-\,2\,\,+ \frac{\tilde{\delta}^2}{\gamma\,
\omega_\Sigma}\Rb^{3/2}} 
 \eeq
 \eq{SOLBE5} leads to the next order corrections for the value of the
 anomalous dimension:
 \beq \label{SOLBE6}
\gamma^{(2)}_{an}\Lb \omega_1, \omega_2\Rb\,\, = \,\,\gamma^{(1)}_{an}\Lb 
\omega_1, \omega_2\Rb\,\,-\,\,\h\,\tilde{\delta}^2\frac{2}{\omega_1 + 
\omega_2}\,\,\,+\,\,\,{\cal O}\Lb \tilde{\delta}^3\Rb
\eeq   
 
  \subsection{Evolution with  Bose-Einstein enhancement:  solution}
    
    The solution to the non-homogenous equation has a general form:
      \bea \label{SOLEN}
 D^{(2)}\Lb Y,Y,\xi\Rb\,\,&=&\,\,\int^{\epsilon + i \infty}_{\epsilon - i \infty}\frac{d \gamma}{ 2 \pi i}
 \int^{\epsilon + i \infty}_{\epsilon - i \infty}\frac{d \omega_\Sigma}{ 2 \pi i} \int^{\epsilon + i \infty}_{\epsilon - i \infty}\frac{d \omega_D}{ 2 \pi i}\,  e^{ 2 \omega_\Sigma Y \,+ \,\gamma \xi}\,\,\\
&\times& \Bigg\{\,\underbrace{\frac{ \,d^{(1)}_{in}\Lb \omega_\Sigma\Rb\,\delta\Lb \omega_D\Rb
}{\Lb \gamma \,-\,\gamma_{an}\Lb \omega_\Sigma,\omega_D \Rb\Rb\,\Lb \gamma \,-\,\gamma_G\Lb \omega_\Sigma\Rb\Rb}}_{\mbox{\small particular solution of non-homogenous equation}}\,\,+\,\,\underbrace{\frac{d^{(2)}_{in}\Lb \omega_1, \omega_2\Rb}{\gamma - \gamma_{an}\Lb \omega_\Sigma, \omega_D\Rb}}_{\mbox{\small general solution of homogenous equation}}\Bigg\}\nn
 \eea

 Taking the  integral over $\gamma$, by closing the  contours  around
the poles:
 $\gamma = \gamma_G\Lb \omega_1 + \omega_2\Rb$ and $ \gamma = 
\gamma_{an}\Lb \omega_\Sigma,\omega_D\Rb$, we obtain
 \bea \label{SOLEN1}
&& D^{(2)}\Lb Y, Y,\xi\Rb\,\,=\,\,\\
&& \bigintsss^{\epsilon + i \infty}_{\!\!\!\!\!\!\!\!\!\!\!\!\!\!\!\!\!\!\!\!\!\epsilon - i \infty}\!\!\!\!\!\frac{d \omega_\Sigma}{ 2 \pi i} \,\, \bigintsss^{\epsilon + i \infty}_{\!\!\!\!\!\!\!\!\!\!\!\!\!\!\!\!\!\!\!\!\!\epsilon - i \infty}\!\!\!\!\!\!\!\frac{d \omega_D}{ 2 \pi i}\,  e^{\omega_\Sigma\, Y \,
 }\,\Bigg\{d^{(2)}_{in}\Lb \omega_\Sigma, \omega_D\Rb  \,e^{ \gamma_{an}\Lb \omega_\Sigma,\omega_D\Rb\xi }
\, +\,\,\frac{d^{(1)}_{in}\Lb \omega_1 + \omega_2\Rb\,\delta\Lb \omega_D\Rb\Bigg(e^{  \gamma_{an}\Lb \omega_\Sigma, \omega_D\Rb \, \xi }\,\,-\,\,e^{ \,\gamma_G\Lb\omega_1+ \omega_2\Rb\,\xi}\Bigg)}{ \gamma_{an}\Lb \omega_\Sigma, \omega_D\Rb \,-\,\gamma_G\Lb \omega_1 + \omega_2\Rb}\Bigg\}\nn
 \eea 
 
 In the DLA both $Y $ and $\xi$ are large, and we can use the
 method of steepest descent  to evaluate the integral. The equations 
for the saddle point
 values for $\omega_1 = \omega^{SP}_1$ and $\omega_2 = \omega^{SP}_2$
 have the form
:
\bea \label{EQSPBE}
Y\,&=&\,\xi  \frac{\partial \gamma_{an}\Lb \omega_\Sigma,\omega_D\Rb}{\partial \omega_1}\,= \xi \Lb \frac{1}{\Lb \omega^{SP}_1\Rb^2} \,+\,  \frac{ \tilde{\delta}}{\Lb \omega^{SP}_1 + \omega^{SP}_2\Rb^2}\Rb;\nn\\
Y\,&=&\,\xi  \frac{\partial \gamma_{an}\Lb \omega_\Sigma,\omega_D\Rb}{\partial \omega_2}\,= \xi \Lb \frac{1}{\Lb \omega^{SP}_2\Rb^2} \,+\,\frac{ \tilde{\delta}}{\Lb \omega^{SP}_1 + \omega^{SP}_2\Rb^2}\Rb;
\eea
 From \eq{EQSPBE}, one can see that $\omega^{SP}_D =0$. Therefore
 \eq{EQSPBE} reduces to
   \beq \label{SOLEN4}
 2 \,Y \,\,=\,\,  2 \frac{\Lb 1 + \h\tilde{\delta}\Rb\,\xi}{\Lb\omega_\Sigma^{SP}\Rb^2};~~~~~~~~~~~~~\omega_\Sigma^{SP}\,\,\sqrt{\frac{\Lb1 +\h \tilde{\delta}\Rb\,\xi}{ \,Y}};  
\eeq   

Integrating over $\omega_D$ expanding $\gamma_{an}\Lb \omega_\Sigma,
\omega_D\Rb\,=\,\gamma_{an}\Lb \omega_\Sigma,\omega_D = 0\Rb\,\,+\,
\,\omega^2_D/\omega^3_\Sigma$ and
 neglecting the contributions that are proportional to $\delta$ 
everywhere,
  except  for the anomalous dimension in the exponent, we have the 
solution in the form:
 \bea \label{SOLEN6}
&& D^{(2)}\Lb  Y,Y,\xi\Rb\,=\,\\
&&\frac{1}{4 \pi}\frac{\Lb\omega^{SP}_\Sigma\Rb^{3/2}}{ \xi}\Big(\frac{4}{3}\,d^{(1)}_{in}\Lb  \omega^{SP}_\Sigma\Rb \,\,+\,\, d^{(2)}\Lb \omega^{SP}_\Sigma,   \omega^{SP}_D=0\Rb\Bigg) e^{4\sqrt{2\, N_c\Lb 1 + \h\tilde{\delta}\Rb \,  Y\,\xi\,}   }\,\,-\,\,\frac{1}{\sqrt{3 \pi}}\sqrt{\frac{\Lb\omega^{SP}_{\Sigma}\Rb^3}{ \xi }}\,\omega^{SP}_{\Sigma}\,d^{(1)}_{in}\Lb \omega^{SP}_{\Sigma} \Rb\,e^{2\, \sqrt{2\,N_c \,  \xi }} \nn
\eea
 Recall, that we  have changed our notation, and $Y = \ln \Lb 
1/x\Rb$ in
   \eq{SOLEN6} .

 The principle difference of the behaviour of the double parton 
density without Bose-Einstein enhancement,  is that we did not 
have a 
term which is proportional to the product of two single parton densities.
 On the other hand,  the violation of \eq{BE1} is small, and we reach a
 value  larger than 10\%, only at $Y\xi \sim \,>\,4$.

At the LHC energies, $Y \sim 16$, and $\xi \geq\,0.25$  appears to be a
 reasonable region for  measurements.

  \section{Non-linear evolution equation for the scattering amplitude}
  \subsection{The  equation}

      We have discussed  the double parton densities, but the cross 
section for
  $J/\psi$ production  depends on the scattering amplitude for
 which the two BFKL Pomerons contribution of the diagram of \fig{fidi}
  and the double gluon density in general, only give a
contribution  in the linear approach.
   The scattering amplitude depends crucially on the shadowing corrections 
(see \cite{KOLEB} for the review). In this section we write the
 generalization of the BK equation\cite{BK,LELU} for the scattering 
 amplitude for two dipoles with a target.
  Following Ref. \cite{BK} the generating functional for the scattering
 amplitude $N$ is defined  
\beq \label{AMP}
 N(Y, [ \gamma_n]) \,\,=\,\,\,\, \sum^{\infty}_{n=2}\,\,(-1)^n\,
\int\,\gamma_n(\rv_1, \bv_1\ldots \,,\rv_n, \bv_n;Y_0)\,\,
\rho^{(n)}(\rv_1\,\bv_1\,\ldots\,,\rv_n, \bv_n; Y\,-\,Y_0)
\,\,\prod^n_{i =1}\,
d^2 r_i \, d^2 b_i \,.
\eeq 
where $\gamma_n$ denotes the amplitude for simultaneous 
scattering of $n$ dipoles off the target, and $\rho^{(n)}$ is the $n$-
 density of the fast projectile, i.e. the two dipoles. As we saw from
 \fig{fidi} at $Y \to Y_0$, the scattering amplitude of  interest
 includes the interaction of two dipoles. From\eq{AMP} one has, that
\beq \label{AMP1}
\rho^{(n)}\,=\,\,(-1)^n\,\frac{\delta\,N}{\delta\,\gamma_n}
\eeq
Using \eq{RON},\eq{EVRON} and \eq{EVRONTI}   we obtain the linear
 functional equation for the amplitude $N$:
\bea \label{AMP3}
\frac{\partial N\Lb Y, [ \gamma_n]\Rb}{ 
\bas\,\partial\,Y}\,\,&=&\,\, \sum^{\infty}_{n=1}\,\,
\int\,\prod^n_{i =1}\,
d^2 \,r_i \, d^2\, b_i \,\,
\gamma_n(\rv_1, \bv_1,\ldots \,,\rv_n, \bv_n;\,Y_0)\,\,\\
&\times&
\,\,\left(
-\,\sum_{i=1}^n
 \,\,\omega(r_i)\,\,\frac{\delta\,N}{\delta\,\gamma_n}\,\,+
2\,\sum_{i=1}^n\,
\int\,\frac{d^2\,r'}{2\,\pi}\,\,
\frac{r'^2}{r^2_i\,(\rv_i\,-\,\rv')^2}\,\,
\frac{\delta\,N}{\delta\,\gamma_n}\,\, 
-\,\,\sum_{i=1}^{n-1}\,\frac{(\rv_i + \rv_n)^2}
{(2\,\pi)\,r^2_i\,r^2_n}\,\,\frac{\delta\,N}{\delta\,\gamma_{n-1}}\,\right)\,.\nn
\eea
The equation (\ref{AMP3}) can be solved using the ansatz:
$N(Y,[\gamma])=\,N(\gamma_1(Y), \gamma_2(Y)\, \ldots )$
and the initial condition that at $Y=Y_0$:
\beq \label{INCONAMP}
\gamma_2\Lb \rv_1,\bv, \rv_2,\bv\Rb \,=\,\gamma_1\Lb \rv_1,
\bv\Rb\,\gamma_1\Lb \rv_2,\bv\Rb\, \delta^{(2)}\Lb \rv - 
\rv_1\Rb\,\delta^{(2)}\Lb \bv - \bv_1\Rb\,\, \delta^{(2)}\Lb
 \rv' - \rv_2\Rb\,\delta^{(2)}\Lb \bv' - \bv_1\Rb;~~\gamma_n \,=
\,0\,~~\mbox{for} \,\,n\neq 2;
\eeq

Using \eq{INCONAMP} we can reduce \eq{AMP3} to the following form
\bea \label{AMP4}
 \frac{\partial \,N^{(2)}\Lb Y, \rv_1,\rv_2,\bv\Rb}{\bas\,\partial\,\,Y}\,\,&=&
 \,\,\sum_{i=1}^2\,\int d^2 r' \frac{r^2_i}{r'^2\,\Lb \rv_i - \rv'\Rb^2} \Bigg\{ 2\,N^{(2)}\Lb Y, \rv' ,\rv_{i+1},\bv- \h(\rv_i - \rv')\Rb  \,\,-\,\,N^{(2)}\Lb Y, \rv_1,\rv_2,\bv\Rb\,\,\nn\\ 
 &-&\,\, N^{(2)}\Lb Y, \rv' ,\rv_{i+1},\bv - \h(\rv_i - \rv')\Rb\,N^{(1)}\Lb Y,\rv_1 -  \rv' , \bv - \h \rv'\Rb\Bigg\}
 \eea
  where $N^{(1)}$ denotes the solution to the Balitsky-Kovchegov equation.
 Note, that in \eq{AMP4} $r_3\equiv  r_1$.

  \subsection{Solution  deep in the saturation region}
 In this section,   using the approach developed in Ref.\cite{LETU},
 we  find the solution to \eq{AMP4},  
deep in the saturation region. 
 In this region both $N^{(2)}$ and $N^{(1)}$ are close to unity, and can
 be written as
 \beq \label{SOLDSR1} 
 N^{(2)}\Lb Y, \rv_1,\rv_2,\bv\Rb\,\,=\,\,1 \,-\,\, \Delta^{(2)}\Lb Y, \rv_1,\rv_2,\bv\Rb;~~~~~~~~~  N^{(1)}\Lb Y, \rv,\bv\Rb\,\,=\,\,1 \,-\,\, \Delta^{(1)}\Lb Y, \rv,\bv\Rb;
 \eeq  
  with $\Delta^{(2)} \,\ll\,\,1$ and $\Delta^{(1)}\,\ll\,1$.  In this
 kinematic region all $\tau_i\,=\,r^2_i\,Q^2_s\Lb Y, b_i\Rb\,\gg\,1$.
   Plugging \eq{SOLDSR1} into \eq{AMP4},  and  neglecting the contribution 
of  
order  $\Delta^{(2)}\,\Delta^{(1)}$, we obtain the following equation for 
$\Delta^{(2)}$.

   \bea \label{SOLDSR2}
   && \frac{\partial \,\Delta^{(2)}\Lb Y, \rv_1,\rv_2,\bv\Rb}{\bas\,\partial\,\,Y}\,\,=\\
 &&\,\,\sum_{i=1}^2\,\int d^2 r' \frac{r^2_i}{r'^2\,\Lb \rv_i - \rv'\Rb^2} \Bigg\{ \Delta^{(2)}\Lb Y, \rv' ,\rv_2,\bv- \h(\rv_i - \rv')\Rb \,\,-\,\,\Delta^{(1)}\Lb Y, \rv' ,\bv- \h(\rv_i - \rv')\Rb \,\,-\,\,\Delta^{(2)}\Lb Y, \rv_1,\rv_2,\bv\Rb\Bigg\}\nn
 \eea   
  First we solve the homogenous equation:
    \bea \label{SOLDSR3}
 \frac{\partial \,\Delta^{(2)}\Lb Y, \rv_1,\rv_2,\bv\Rb}{\bas\,\partial\,\,Y}\,\,&=&
\h\sum_{i=1}^2\int d^2 r' \frac{r^2_i}{r'^2\,\Lb \rv_i - \rv'\Rb^2} \Bigg\{ 2 \Delta^{(2)}\Lb Y, \rv' ,\rv_2,\bv- \h(\rv_i - \rv')\Rb \,\,-\,\,\-\,\,\Delta^{(2)}\Lb Y, \rv_1,\rv_2,\bv\Rb\Bigg\} \,\,\nn\\
&- & \,\,\h( \omega\Lb r_1\Rb + \omega\Lb r_2\Rb) \Delta^{(2)}\Lb Y, \rv_1,\rv_2,\bv\Rb
 \eea   
   Inside  the saturation region\cite{LETU}
 $\omega\Lb r_i\Rb \,\,=\,\,\ln\Lb \tau_i\Rb\,=\,\ln \Lb r^2_i\,Q_s^2\Lb Y, \bv\Rb\Rb\,\,=\,\,z_i$
   and therefore the last term of the l.h.s. of the equation has
 a contribution:$\h
   ( z_1 + z_2) \Delta^{(2)}$. Bearing  in mind that we  are searching
 for the solution of \eq{SOLDSR3}  in the form $\Delta^{(2)}\Lb z_1 + z_2\Rb$.
  To  find the solution we go to the Mellin transform given by
   \beq \label{SOLDSR4}
   \Delta^{(2)}\Lb z_1 + z_2\Rb \,\,=\,\,\int^{\epsilon + i \infty}_{\epsilon
 - i \infty}\,\frac{d \gamma}{2 \pi i }  e^{\gamma \Lb z_1 + z_2\Rb} 
\,\widetilde{\Delta}^{(2)}\Lb \gamma\Rb
   \eeq
  The term in $\{\dots\}$ in \eq{SOLDSR3}  gives a contribution which
 is equal to $\h \chi\Lb \gamma \Rb  \widetilde{\Delta}^{(2)}\Lb 
\gamma\Rb$,
 where $\chi$ is the BFKL kernel\cite{BFKL,KOLEB}:
  \beq \label{BFKLKER}
 \chi\Lb \gamma\Rb \,=\,2 \psi\Lb 1\Rb - \psi\Lb \gamma\Rb - \psi\Lb 1 -
 \gamma\Rb;
\eeq
 where $\psi(x) = d \ln \Gamma(x)/d x$  and $\Gamma$ is the Euler gamma
 function \cite{RY}. Taking into account that $\ln\Lb Q^2_s\Lb Y, 
\bv\Rb\Rb\,=\,\kappa\,Y $ with $\kappa \,=\,\,\frac{\chi\Lb 1
 - \gamma_{cr}\Rb}{1 - \gamma_{cr}} $, where $\gamma_{cr}$  in
 the leading log approximation, which we use in this paper,
 stems from the solution of the equation:
 \beq \label{GA}
\frac{\chi\Lb \gamma_{cr}\Rb}{1 -   \gamma_{cr}}\,\,=\,\Big{|}
 \frac{d \chi\Lb \gamma_{cr}\Rb}{d \gamma_{cr}}\Big{|}
\eeq
and     $ \gamma_{cr}\,\approx\,\,0.37$.

Finally,  \eq{SOLDSR3}   takes the form:
\beq \label{SOLDSR5}
2 \kappa\, \gamma\,\widetilde{\Delta}^{(2)}\Lb \gamma\Rb\,=\,\chi\Lb \gamma\Rb \,\widetilde{\Delta}^{(2)}\Lb \gamma\Rb\,\,
+\,\,\h\frac{d \widetilde{\Delta}^{(2)}\Lb \gamma\Rb}{d \gamma}
\eeq   
Solving this equation we obtain
\beq \label{SOLDSR6}
\widetilde{\Delta}^{(2)}\Lb \gamma\Rb\,\,=\,{\rm C}\,\exp\Lb 2 
\kappa \,\gamma^2 \,-\,2\,\int^\gamma\!\!\!\! d \gamma' \,\chi\Lb \gamma'
 \Rb\Rb
\eeq
where ${\rm C}$ is the arbitrary constant. From \eq{SOLDSR4} we obtain that
\beq \label{SOLDSR7}
  \Delta^{(2)}\Lb z_1 + z_2\Rb\,\,=\,\,\,{\rm C}\,\int^{\epsilon + i \infty}_{\epsilon - i \infty}\,\frac{d \gamma}{2 \pi i }  \exp\Lb \gamma \Lb z_1 + z_2\Rb \,+\,2\,\kappa \,\gamma^2 \,-\,2\,\int^\gamma\!\!\! d \gamma' \,\chi\Lb \gamma' \Rb\Rb
  \eeq  
  
  Taking the integral over $\gamma$ using the method of steepest descent, 
we obtain the following equations for the saddle point $\gamma^{SP}$ 
  \beq \label{SOLDSR8}   
  (z_1 + z_2)\,+\,4 \kappa \gamma^{SP}  -  2\,\chi\Lb \gamma^{SP}\Rb\,=\,0
  \eeq
  
  For large values of $z_1 + z_2$ \,$\gamma^{SP} \,\,=\,\,-\frac{1}
{4\,\kappa}\Lb \Lb z_1 + z_2\Rb\,+\,2\,\chi\Lb - \frac{1}{2\,
\kappa}(z_1+z_2)\Rb\Rb$ since $\chi(\gamma) \xrightarrow{ \gamma \gg 1}
  \ln(\gamma)$. 
    
  Therefore,
    \beq \label{SOLDSR8}  
      \Delta^{(2)}\Lb z_1 + z_2\Rb\,\,=\,\,\,{\rm C}\exp\Lb - \frac{\Lb z_1 
+ z_2\Rb^2}{8 \kappa} \,+\,\, \frac{1}{2 \kappa} \chi^2\Lb - \frac{1}
{2\,\kappa}(z_1+z_2)\Rb\,\,-\,\,\int^{-\frac{1}{2\,\kappa}\, \Lb z_1 +
 z_2\Rb}\!\!\! d \gamma' \,\chi\Lb \gamma' \Rb \Rb \eeq
  Since 
  \beq \label{SOLDSR9}
  \int^\gamma\!\!\! d \gamma' \,\chi\Lb \gamma' \Rb\,\,=\,\,-2\,\gamma_E \,\gamma  \,+\,\ln\Lb \frac{\Gamma\Lb 1 - \gamma\Rb}{\Gamma\Lb \gamma\Rb}\Rb\,\,\xrightarrow{\gamma \gg 1}  \,\,- 2 \gamma \ln(\gamma)
  \eeq
    at large $z_1 + z_2$ we see that
     \beq \label{SOLDSR10}  
   \Delta^{(2)}_{ h.\,\,eq.}\Lb z_1 + z_2\Rb\,\xrightarrow{z_1 + z_2 \gg
 1}\,\,{\rm C} \,\,e^{ - \frac{\Lb z_1 + z_2\Rb^2}{8 \,\kappa} \, 
 -\,\frac{\Lb z_1 + z_2\Rb}{2 \,\kappa} \,\ln\Lb z_1 + z_2\Rb}  
   \eeq
   where  {\it h.eq.} denotes  homogenous equation.
   
   It should be noted that the main contribution to the large $z_1
 + z_2$ asymptotic behavior, stems from the
   term: $\h( \omega\Lb r_1\Rb + \omega\Lb r_2\Rb) 
\Delta^{(2)}\Lb Y, \rv_1,\rv_2,\bv\Rb $ in the homogenous
 equation (see \eq{SOLDSR3}), which is responsible for 
the gluon reggeization. Having this in mind, we can
 simplify the non-homogenous \eq{SOLDSR2}, replacing it by
    \beq \label{SOLDSR11}
    \frac{\partial \,\Delta^{(2)}\Lb Y, \rv_1,\rv_2,\bv\Rb}{\bas\,\partial\,\,Y}\,\,=\\
\,\h\Lb \omega(r_1) + \omega(r_2)\Rb  \Delta^{(2)}\Lb Y, \rv'_1,\rv_2,\bv\Rb \,\,-
\,\,\h \omega(r_1) \Delta^{(1)}\Lb Y, \rv_1 ,\bv\Rb \,\,-\,\,\h \omega(r_2) \Delta^{(1)}\Lb Y, \rv_2 ,\bv\Rb \eeq    
   Introducing new variables: $z_{12} = z_1 + z_2$ and $\zeta_{12} = z_1
 - z_2$, we ca re-write \eq{SOLDSR11}        
in the form:
   \beq \label{SOLDSR12}
  2 \kappa\,  \frac{\partial \,\Delta^{(2)}\Lb z_{12},\zeta_{12}\Rb}{\partial\,\,z_{12}}\,\,=
\,\h\,z_{12} \Delta^{(2)}\Lb z_{12},\zeta_{12} \Rb \,\,-
\,\,\frac{1}{4}\Lb z_{12} + \zeta_{12}\Rb  \Delta^{(1)}\Lb z_{12} + \zeta_{12}\Rb \,\,-\,\, \frac{1}{4}\Lb z_{12} - \zeta_{12}\Rb  \Delta^{(1)}\Lb z_{12} - \zeta_{12}\Rb\eeq
The  solution to this equation has a general form
\bea \label{SOLDSR13}
\Delta^{(2)}\Lb z_{12},\zeta_{12} \Rb\,\,&=&\,\,\Delta^{(2)}_{h.\,\,eq.}\Lb z_{12},\zeta_{12} \Rb\\
&-&\,\,\frac{1}{4}\Delta^{(2)}_{h.\,\,eq.}\Lb z_{12},\zeta_{12} \Rb\,\int^{z_{12}}_0 d z'_{12}\frac{\Bigg(\Lb z'_{12} + \zeta_{12}\Rb  \Delta^{(1)}\Lb z'_{12} + \zeta_{12}\Rb \,\,+\,\, \Lb z'_{12} - \zeta_{12}\Rb  \Delta^{(1)}\Lb z'_{12} - \zeta_{12}\Rb\Bigg)}{\Delta^{(2)}_{h.\,\,eq.}\Lb z'_{12},\zeta_{12}\Rb}\nn
\eea

Plugging in \eq{SOLDSR13} $\Delta^{(1)}\Lb z_{12}+ \zeta_{12} \Rb\,
\,=\,\,{\rm C^{(1)}}\exp\Lb -\Lb z_{12}+ \zeta_{12} \Rb/(8\,
\kappa)\Rb$\cite{LETU} we obtain the general solution in the following form:
\bea \label{SOLDSR14}
\Delta^{(2)}\Lb z_{12},\zeta_{12} \Rb\,\,&=&\,\,{ \rm C}^{(2)}\Lb
 \zeta_{12}\Rb\,\Delta^{(2)}_{h.\,\,eq.}\Lb z_{12}\Rb \,  -\, 
\kappa \Bigg(  \Delta^{(1)}\Lb z'_{12} + \zeta_{12}\Rb \,\,+\,\, 
  \Delta^{(1)}\Lb z'_{12} - \zeta_{12}\Rb\Bigg)
\nn\\
&=& \,{ \rm C}^{(2)}\Lb \zeta_{12}\Rb\,\exp\Lb -z^2_{12}/(8\,\kappa)\Rb -
 \kappa\,{\rm C^{(1)}}\Bigg(\exp\Lb  - z^2_1/(2\,\kappa)\Rb\,+\, \exp\Lb  - z^2_2/(2\,\kappa)\Rb\Bigg)\eea
 where ${\rm C^{(2)}}\Lb z_1 - z_2\Rb$ is an arbitrary
 function of $ z_1 - 
z_2$, which should be found from the initial conditions.

One can see that the scattering amplitude shows  geometric scaling
 behavior depending on two variable $z_1$ and $z_2$, instead of three 
variables: $Y,r_1,r_2$. For $z_1 \approx  z_2$ $\Delta^{(2)} \propto 
\Delta^{(1)}\Lb z_1 \approx z_2\Rb$ while for $z_1 \,\gg\,z_2\, \gg \,1$
 we see that $\Delta^{(2)}\Lb z_1\Rb\, \propto\, \exp\Lb - z^2_1/(8 
\kappa)\Rb \,\gg\,\Delta^{(1)}\Lb z_1 \Rb$.

  \subsection{Solution in the vicinity of the saturation scale.}
  As  has been discussed in Refs.\cite{GLR,MUT}, one does not need
 to know the exact form of the shadowing corrections to find
 the behaviour of the scattering amplitude ($N^{(1)}\Lb Y,r,b\Rb$),
  in the vicinity of the saturation scale. One should find the
 solution to the linear equation and the equation for the saturation
 momentum has the form: $N\Lb 
  Y, r^2 = 1/Q^2_s(Y, b), b\Rb = N_0 \,<\,1$, where $N_0$ is an arbitrary
 constant, which is numerically small ( say $N_0 =1/3$).
  
  We can find the solution to the linear equation  using the 
Mellin transform:
   \beq \label{SOLVSS1}
 N^{(2)}\Lb Y,\xi_1, Y, \xi_2; b\Rb\,\,=\,\,\int^{\epsilon + i
 \infty}_{\epsilon - i \infty}\frac{d \gamma_1}{2 \pi i}
\int^{\epsilon + i \infty}_{\epsilon - i \infty}\frac{d 
\gamma_2}{2 \pi i}\int^{\epsilon + i \infty}_{\epsilon - i
 \infty}\frac{d \omega}{2 \pi i}\,\,e^{ (1 -  \gamma_1) \,\xi_1 \,+\,(1 -  \gamma_2)\,xi_2 \, +\, \omega\, Y}\,d^{(2)}\Lb \omega,\gamma_1,\gamma_2\Rb\eeq
where $\xi_i \,\,=\,\,\ln\Lb r^2_i \mu^2\Rb$ where $\mu$ is the soft scale.

For $d^{(2)}\Lb \omega,\gamma_1,\gamma_2\Rb$ the equation takes the form:
\beq 
\omega\,d^{(2)}\Lb \omega,\gamma_1,\gamma_2\Rb\,=\,\,\bas\,
\Big\{ \chi\Lb \gamma_1\Rb \,+\,\chi\Lb \gamma_2\Rb\Big\} d^{(2)}\Lb \omega,\gamma_1,\gamma_2\Rb
\eeq 

Therefore, the amplitude $ N^{(2)}\Lb Y,\xi_1, Y, \xi_2; b\Rb$ is equal
 to
  \beq   \label{SOLVSS2}
  N^{(2)}\Lb Y,\xi_1, Y, \xi_2; b\Rb\,=\,\int^{\epsilon + i \infty}_{\epsilon - i \infty}\frac{d \gamma_1}{2 \pi i}
\int^{\epsilon + i \infty}_{\epsilon - i \infty}\frac{d \gamma_2}{2 \pi i}\, d^{(2)}_{\rm in}\Lb\gamma_1,\gamma_2\Rb\,\exp\Bigg( 
 \Big\{ \chi\Lb \gamma_1\Rb \,+\,\chi\Lb \gamma_2\Rb\Big\}\,Y\,+\,(1-   \gamma_1) \,\xi_1 \,+\, (1 - \gamma_2)\,\xi_2 \Bigg)
 \eeq 
 
\begin{figure}[ht]
    \centering
  \leavevmode
  \begin{tabular}{c c}
      \includegraphics[width=8.2cm]{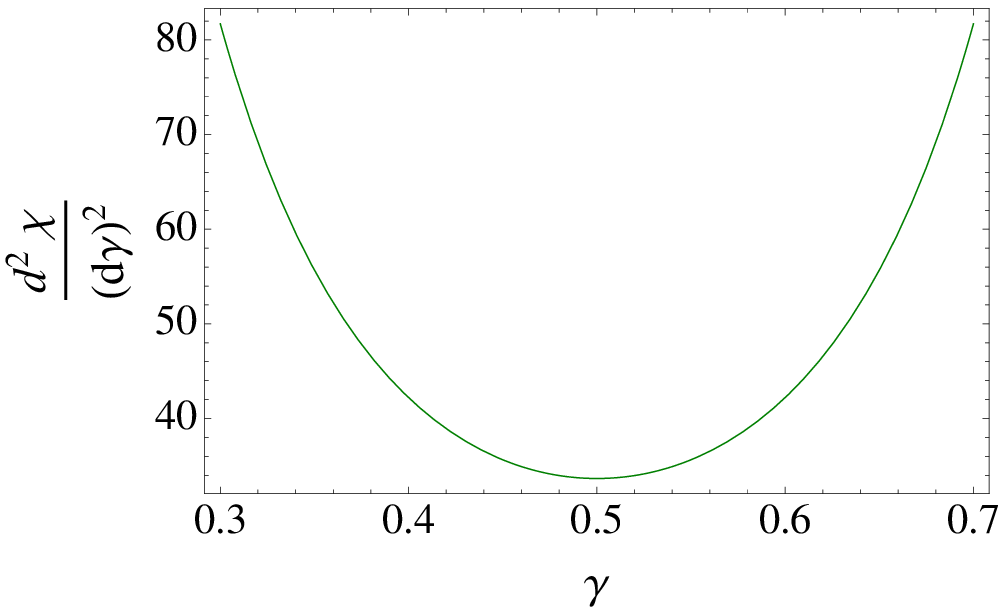} & \includegraphics[width=7cm,height=5cm]{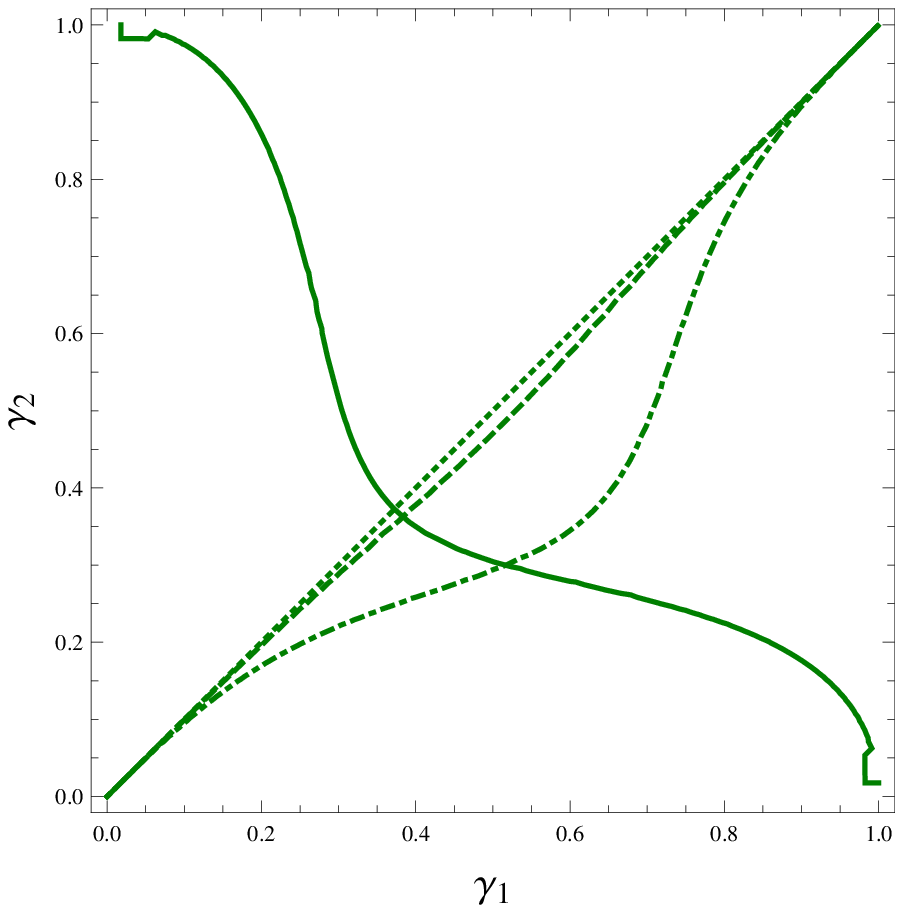}\\
      \fig{ga}-a & \fig{ga}-b\\
      \end{tabular}    \caption{$d^2 \chi(\gamma)/(d \gamma)^2$
 (\fig{ga}-a) and the solution to \eq{SOLVSS4} (\fig{ga}-b). In
 \fig{ga}-b dotted, dashed and dot-dashed lines corresponds to
 $(\xi_1 - \xi_2)/Y = 0, 1,10$, respectively. The solid line corresponds to the term $ \Big\{ \chi\Lb \gamma^{\rm SP}_1\Rb \,+\,\chi\Lb \gamma^{\rm SP}_2\Rb\Big\}\,$ in \eq{SOLVSS4}}
\label{ga}
  \end{figure}

 Taking the integral using the method of steepest descent, we obtain
 the following equations for the saddle points $\gamma^{\rm \small SP}_1$
 and $\gamma^{\rm \small SP}_2$ :
 \beq \label{SOLVSS3}
  \frac{d\,\chi\Lb \gamma^{\rm SP}_1\Rb}{d \,\gamma^{\rm \small SP}_1}\,Y
\,-\,\xi_1\,\,=\,\,0;~~~~~~~~~~~  
   \frac{d\,\chi\Lb \gamma^{\rm \small SP}_2\Rb}{d \,\gamma^{\rm \small SP}_2}\,Y\,-\,\xi_2\,\,=\,\,0; 
   \eeq
   The line, on which $  N^{(2)}\Lb Y,\xi_1, Y, \xi_2; b\Rb$ is  constant, 
 is given by the equation:
    \beq \label{SOLVSS4}    
    \Big\{ \chi\Lb \gamma^{\rm SP}_1\Rb \,+\,\chi\Lb \gamma^{\rm SP}_2\Rb\Big\}\,Y\,+\,(1-   \gamma^{\rm SP}_1) \,\xi_1 \,+\, (1 - \gamma^{\rm SP}_2)\,\xi_2 \,=\,0
    \eeq   
    One can see that the difference $\Delta \gamma^{\rm SP} = 
 \gamma^{\rm SP}_1\,-\,\gamma^{\rm SP}_2$ turns out to be small. Indeed,
    \beq \label{SOLVSS5} 
    \Bigg(  \frac{d\,\chi\Lb \gamma^{\rm SP}_1\Rb}{d \,\gamma^{\rm  SP}_1}\,\,-\,\,    \frac{d\,\chi\Lb \gamma^{\rm SP}_2\Rb}{d \,\gamma^{\rm SP}_2}\,\Bigg) \,Y\,  =\,\,\frac{d^2 \chi\Lb \h(\gamma^{\rm SP}_1 +  \gamma^{\rm SP}_2)\Rb}{\Lb d(\h(\gamma^{\rm SP}_1 +  \gamma^{\rm SP}_2)\Rb^2}\,Y\,\Delta \gamma^{\rm SP}  \,\,=\,\xi_1 - \xi_2
   \eeq    
   In \fig{ga}-a we show that $d^2 \chi(\gamma)/d \gamma^2$
 turns out to be large, as well as the value of $Y$. Therefore, for
 reasonable values of $\xi_1 - \xi_1 \leq Y$,    \eq{SOLVSS5} leads to a
large  value of $\Delta \gamma^{\rm SP}$.
 Bearing this in mind we
 see that the solution of \eq{SOLVSS4} coincides with   
 the solution of \eq{GA}.
  \fig{ga}-b gives the solution of \eq{SOLVSS4} when 
$\gamma^{\rm SP}_1 \neq \gamma^{\rm SP}_2$. One can see that a 
significant difference for $\Delta \gamma^{\rm SP}$ can be
 obtained for $(\xi_1 - \xi_2)/Y \approx 10$.
  
  Generally, the solution to \eq{SOLVSS2}  has the following
 dependence on $r_1$ and $r_2$:
  \beq \label{SOLVSS6}  
   N^{(2)}\Lb Y, r_1, Y, r_2; b\Rb \,\,=\,\,{\rm Const}\,\Lb r^2_1 \,Q^2_1\Lb Y, b\Rb\Rb^{1 - \gamma^{\rm cr}_1}\,
   \Lb r^2_2 \,Q^2_2\Lb Y, b\Rb\Rb^{1 - \gamma^{\rm cr}_2}\,  
   \eeq
   where $\gamma^{\rm cr}_1$ and $ \gamma^{\rm cr}_2$ are 
solutions to \eq{SOLVSS4} (see \fig{ga}-b) and   
   \beq \label{SOLVSS7}
   \ln\Lb Q^2_1\Lb Y, b\Rb/Q^2_1\Lb Y = Y_0 , b\Rb \Rb\,\,=\,\,\frac{\chi\Lb \gamma^{\rm cr}_1\Rb}{1 - \gamma^{\rm cr}_1}\,Y;
   ~~~~~~~~~~~~~~~~~  \ln\Lb Q^2_2\Lb Y, b\Rb/Q^2_1\Lb Y = Y_0 , b\Rb \Rb\,\,=\,\,\frac{\chi\Lb \gamma^{\rm cr}_2\Rb}{1 - \gamma^{\rm cr}_2}\,Y;   
     \eeq

  \subsection{Initial conditions}
  Taking  the shadowing corrections into account for the scattering
 amplitude, we can re-write  \eq{FD9} for the cross section
 of $J/\psi$ production in the form:
  \beq \label{IC1}
 \sigma\Lb Y, Q^2\Rb\,\,= \,\,4 \bas^2 \,\, \int d^2 r\,d^2 r'\,\Phi\Lb \vec{r},\vec{r}'\Rb \int d^2 b\,  N^{(2)}_A\Lb x, \vec{r} ; x, \vec{r}' ;  \vec{b}_T\Rb 
 \eeq  
  From \eq{FD2} and \eq{FD10} we find that \eq{IC1} takes the following 
form:
  \bea \label{IC2}
   \sigma\Lb Y, Q^2\Rb\,\,&=& \,\,4 \bas^2 \,\,\int \frac{d^2 r_1}{4 \pi} \,\, \Psi_{\gamma^*}\Lb Q , r _1\Rb \,\Psi_{J/\psi}\Lb r _1\Rb \,  \,\int \frac{d^2 r_2}{4 \pi} \,, \Psi_{\gamma^*}\Lb Q , r _2\Rb \,\Psi_{J/\psi}\Lb r _2\Rb \nn\\
   &\times&\,\Bigg\{N^{(2)}_A\Lb Y, \h(\rv_1 - \rv_2) ; Y, \h(\rv_1 - \rv_2) ;  \vec{b}_T\Rb \,-\,N^{(2)}_A\Lb Y, \h(\rv_1 + \rv_2) ; Y, \h(\rv_1 - \rv_2) ;  \vec{b}_T\Rb\nn\\ \,
   &-&\,  N^{(2)}_A\Lb Y, \h(\rv_1 + \rv_2) ; Y, \h(\rv_1 - \rv_2) ;  \vec{b}_T\Rb  \,+\,   N^{(2)}_A\Lb Y, \h(\rv_1 - \rv_2) ; Y, -\h(\rv_1 - \rv_2) ;  \vec{b}_T\Rb  \Bigg\}\nn\\
   &=&4 \bas^2 \,\,\int \frac{d^2 r_1}{4 \pi} \, \Psi_{\gamma^*}\Lb Q , r _1\Rb \,\Psi_{J/\psi}\Lb r _1\Rb \,  \,\int \frac{d^2 r_1}{ 4 \pi} \,\Psi_{\gamma^*}\Lb Q , r _2\Rb \,\Psi_{J/\psi}\Lb r_2\Rb\nn\\
    &\times&\,2\Bigg\{N^{(2)}_A\Lb Y, \h(\rv_1 - \rv_2) ; Y, \h(\rv_1 - \rv_2) ;  \vec{b}_T\Rb \,-\,N^{(2)}_A\Lb Y, \h(\rv_1 + \rv_2) ; Y, \h(\rv_1 - \rv_2) ;  \vec{b}_T\Rb\Bigg\}     \eea
    
     To solve \eq{AMP4}, we need to put in the initial conditions.
 First, we
 need to find the solution to the BK equation, and the initial conditions
 for $N^{(1)}\Lb Y,  r,  b\Rb$ for  dipole scattering with nuclei, this 
is 
given by the McLerran-Venugopalan formula\cite{MV}
  \beq \label{IC3}
  N^{(1)}\Lb Y = Y_0 ,  r,  b\Rb  \,\,=\,\,1 \,-\,\exp\Lb - \sigma_{\rm dn}\Lb r\Rb  S_A\Lb b \Rb\Rb \xrightarrow{ \rm QCA}\, 1 \,-\,\exp\Lb - \frac{1}{8} r^2 Q^2_s\Lb Y_0,b\Rb\Rb
  \eeq
  where the dipole-nucleon cross section ($\sigma_{\rm dn}$) 
 in the Born approximation of perturbative QCD is equal to 
    $ \sigma_{\rm dn}\,=\,4 \pi \as^2 r^2 \,\ln(1/(r^2\,\mu^2)$ where
 $ \mu $ is the soft scale.  In the quasi-classical approach (QCA) we
 replace this cross section by $\sigma_{\rm dn}\,=\frac{1}{8} r^2\,Q^2_s\Lb
 Y=Y_0,b\Rb$.
    
     The initial condition for    $\Big\{ \dots \Big\}  $  in \eq{IC2} can 
be found using the approach of Refs.\cite{KLNT,DKLMT}.  For $J/\psi$
 production at low energy, the $q \bar{q}$ pair  propagate from first
 to the last inelastic interaction (see \fig{ba}-a points $z_1$ and
 $z_2$, respectively), undergoing  inelastic interactions  with the
 cross section $\frac{1}{16}\Lb \rv_1 - \rv_2\Rb^2\,Q^2_s\Lb Y_0,b\Rb$.
 Before the first inelastic collisions and after the last one, the
 quark-antiquark pair has only elastic rescatterings. Plugging in
 the expression for the first and the last inelastic interaction,
 given in Refs.\cite{KLNT,DKLMT},    we obtain
     \bea \label{IC4}
     &&\Bigg\{N^{(2)}_A\Lb Y, \h(\rv_1 - \rv_2) ; Y, \h(\rv_1 - \rv_2) ;  \vec{b}_T\Rb \,-\,N^{(2)}_A\Lb Y, \h(\rv_1 + \rv_2) ; Y, \h(\rv_1 - \rv_2) ;  \vec{b}_T\Rb\Bigg\}_{Y=Y_0}  \,\,=\nn\\
     &&\,\,Q^2_s\Lb Y=Y_0, b\Rb \frac{\Lb \rv_1\cdot\rv_2\Rb^2}{\Lb \rv_1 + \rv_2\Rb^2}\,\Bigg(   \exp\Lb - \frac{1}{16}\,Q^2_s\Lb Y_0,b\Rb \Lb \rv_1 - \rv_2\Rb^2\Rb\,\,-\,\,   \exp\Lb - \frac{1}{8}\,Q^2_s\Lb Y_0,b\Rb \Lb  r^2_1 + r^2_2\Rb\Rb  \Bigg)
     \eea 
 In Ref.\cite{KMV}( see also Refs.\cite{DMXY,BGV})  it is shown
 how \eq{IC4} stems from the scattering amplitude of two dipoles
 with the nucleus target in the large $N_c$ limit.
            \begin{boldmath}
  \subsection{ Wave functions for virtual photon and $J/\psi$}
  \end{boldmath}
       For the completeness of presentation we need to specify what 
expression we use for the wave functions in \eq{AMP4}. 
       We adopt the wave functions from Refs.\cite{KOTE,MPS} which have
 the following form,
      Denoting $ \Psi_{\gamma^*}\Lb Q , r _1\Rb \,\Psi_{J/\psi}\Lb r _1 \Rb = \int d z \,\Phi_{L,T}^{\gamma^*J/\psi}(z,\rv,Q^2) $      
       \begin{eqnarray}\label{WF1}
\Phi^{\gamma^*J/\psi}_L(z,\rv,Q^2) 
   & = & \hat{e}_f \sqrt{\frac{\alpha_e}{4\pi}} N_c \, 2QK_0(r\bar Q_f)  M_{J/\psi}z(1-z)\phi_L(r,z),\nn\\
\Phi^{\gamma^* J/\psi}_T(z,\rv,Q^2) 
   & = & \hat{e}_f \sqrt{\frac{\alpha_e}{4\pi}} N_c\frac{\alpha_e 
N_c}{2\pi^2}\left\{
             m_f^2 K_0(r\bar Q_f)\phi_T(r,z)
           - [z^2+(1-z)^2]\bar Q_f K_1(r\bar Q_f) \partial_r\phi_T(r,z)
            \right\},
\end{eqnarray}       
where $e_f$ denotes the charge of the $c$-quark and $m_c$  is its mass.
 $\bar{Q}^2_f \,=\,m^2_c z (1-z) + Q^2$. $L$ and $T$ denote the
 polarizations of the photon.
For $\phi_{L,T}T(r,z)$ we suggest to use the simplest Gaussian parameterization.

\beq \label{WF2}
\phi_L \, = \, N_L \,\exp\left[-r^2/(2R_L^2)\right];~~~~~~~~~~~~~~~~
\phi_T \, = \, N_T\,z(1-z)\,\exp\left[-r^2/(2R_T^2)\right];
\eeq     

All numerical parameters can be found in Refs.\cite{KOTE,MPS}.

~

~

  \section{Bose-Einstein enhancement and shadowing}
\begin{figure}[ht]
    \centering
  \leavevmode
      \includegraphics[width=10cm]{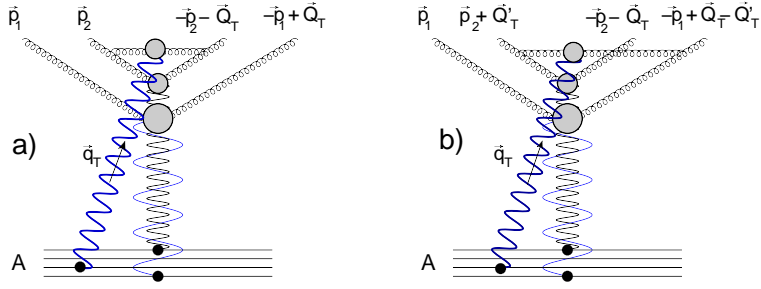} 
    \caption{ The exchange of the extra BFKL Pomeron  for
 DGLAP (\fig{indisc}-a) and interference (\fig{indisc}-b)
  diagrams in DLA of perturbative QCD. \fig{indisc}-a and
 \fig{indisc}-b show the emission an extra gluon in one of
 the parton showers and its interaction with the target due
 to the BFKL Pomeron exchange. }
\label{indisc}
  \end{figure}

At first sight, the increase of the double parton density, which is
 stronger than the product of two single parton densities, is
 potentially harmful, and could lead to a violation of the
 unitarity constraints in the framework of the BFKL Pomeron
 calculus. On the other hand the
JIMWLK equation\cite{JIMWLK} which includes all corrections
 of the order of $1/(N^2_c - 1)$ does not exhibit any problem with 
unitarity.
The goal of this section is to show that it is sufficient  that
 $N^{(1)} \to 1 - \Delta^{(1)}$ for large $Y$, with $\exp\Lb
 \Delta_\pom Y\Rb \,
\Delta^{(1)}\Lb  Y, \rv_1,\rv_2,\bv\Rb\,\ll\,1$, to ensure 
 unitarity at high energies. The reason for this feature can
 be illustrated using \eq{AMP4} in which we insert $N^{(1)} \,=\,1$.
 Such an equation determines the asymptotic solution for $N^{(2)}$. 
The resulting equation takes the form:
    \bea \label{BESC1}
 \frac{\partial \,N^{(2)}\Lb Y, \rv_1,\rv_2,\bv\Rb}{\bas\,\partial\,\,Y}\,\,&=& \,\,\sum_{i=1}^2\,\int d^2 r' \frac{r^2_i}{r'^2\,\Lb \rv_i - \rv'\Rb^2} \Bigg\{ \,N^{(2)}\Lb Y, \rv' ,\rv_{i+1},\bv- \h(\rv_i - \rv')\Rb  \,\,-\,\,N^{(2)}\Lb Y, \rv_1,\rv_2,\bv\Rb\Bigg\}\nn\\
  &=& \h \sum_{i=1}^2\,\int d^2 r' \frac{r^2_i}{r'^2\,\Lb \rv_i - \rv'\Rb^2} \Bigg\{ \,2N^{(2)}\Lb Y, \rv' ,\rv_{i+1},\bv- \h(\rv_i - \rv')\Rb  \,\,-\,\,N^{(2)}\Lb Y, \rv_1,\rv_2,\bv\Rb\Bigg\}\nn\\
   & \,-\,& \,\h\bas\,\Lb \omega(r_1)+ \omega(r_2)\Rb \,N^{(2)}\Lb Y, \rv_1,\rv_2,\bv\Rb  
  \eea    
    The first line of this equation shows that $N^{(2)} = {\rm Const}$ is
 the solution of the equation. The second line illustrates that $N^{(2)}$
 approaches the asymptotic solution with the function
 $\Delta^{(2)}\Lb Y, \rv_1,\rv_2,\bv\Rb \,\propto\,\,
 \exp\Lb -\,\h\bas \Lb \omega(r_1)+ \omega(r_2)\,Y\Rb\Rb $ ($N^{(2)}
 = {\rm Const} - \Delta^{(2)}$).  It should be stressed that
 the fact that $N^{(2)} $ approaches a constant, does not depend 
on the value of $\bas$ which determines the intercept of  the energy
 ($Y$) dependence of the solution to the linear equation. This screening
 is provided by the elastic amplitude, which is equal to unity at high
 energy. Hence, the elastic  rescattering of patrons 
creates a  survival probability\cite{SPBJ,GLMSP}, which suppresses the
 power-like increase of the double parton amplitudes.
We wish to show that $1/(N^2_c - 1)$  corrections to  the
 evolution equation 
have 
the same structure as \eq{BESC1}.  For the sake of simplicity, we
 discuss the shadowing corrections in the framework of the DLA of
 perturbation QCD (see \eq{EQBE2} and \fig{dglapf}). \fig{indisc}
 illustrates that the emission of an extra gluon can be screened
 by the exchange of the BFKL Pomeron, both for normal DGLAP evolution,
 as well as for the interference diagram.

The advantage of the evolution equation (see \eq{AMP4}), is that we can
 take into account the shadowing corrections in the same
way for both terms of the equation for the double gluon structure
 function. The homogenous part of the resulting equation takes
 the form:
\bea \label{BESC2}
\frac{\partial {\cal N}^{(2)}\Lb Y, Y,\xi\Rb}{\partial \,\xi} \,\,&=&\,\, \int^{Y}_{Y_0}\!\!\!\! d Y' \,\,{\cal N}^{(2)}\Lb Y' ,Y,\xi\Rb\Bigg\{1\,\,-\,\,\frac{\bas}{\pi R^2_A \,Q^2} \,{\cal N}^{(1)}\Lb Y', \xi\Rb\Bigg\}\nn\\
&+&\,\, 
\int^{Y}_{Y_0} \!\!\!\! d Y'\,\,{\cal N}^{(2)}\Lb Y, Y', \xi\Rb\,\Bigg\{1\,\,-\,\,\frac{\bas}{\pi R^2_A \,Q^2} \,{\cal N}^{(1)}\Lb Y', \xi\Rb\Bigg\}\,\nn\\
&+&\,\,\frac{2}{N^2_c - 1}
\int^Y\!\!\!\!  d Y'\,{\cal N}^{(2)}\Lb Y', Y',\xi\Rb \Bigg\{1\,\,-\,\,\frac{\bas}{\pi R^2_A \,Q^2} \,{\cal N}^{(1)}\Lb Y', \xi\Rb\Bigg\}
 \eea   
where 
\beq \label{BESC3}
\frac{1}{\pi R^2_A}\,\,\equiv\,\,\int d^2 q_T \,S_A\Lb q_T\Rb
\eeq
${\cal N}^{(2)}\Lb Y, Y,\xi\Rb$ denotes the scattering amplitude, which 
 in the linear approximation coincides with the double parton distribution
 function. However, this amplitude, as can be seen from \eq{BESC2} and
 \fig{indisc},  contains  shadowing corrections, while 
the double parton distribution function cannot be affected by the
 shadowing, as we have seen above.  This screening is provided by
 the elastic amplitude which is equal to unity at high energy.
 Hence, the elastic  rescattering of patrons 
creates a  survival probability\cite{SPBJ,GLMSP}, which suppress
 the power-like increase of the double parton amplitudes\eq{BESC2}, and
 has a typical form for the shadowing correction in DLA of
 perturbative QCD\cite{GLR,MUQI,KOLEB}. However, to see all numerical
 coefficients and to discuss the influence of  shadowing on
 the asymptotic behavior of the scattering amplitude, it is
 desirable to re-write \eq{BESC2} in the coordinate representation. 

\begin{figure}[ht]
    \centering
  \leavevmode
      \includegraphics[width=14cm,height=3cm]{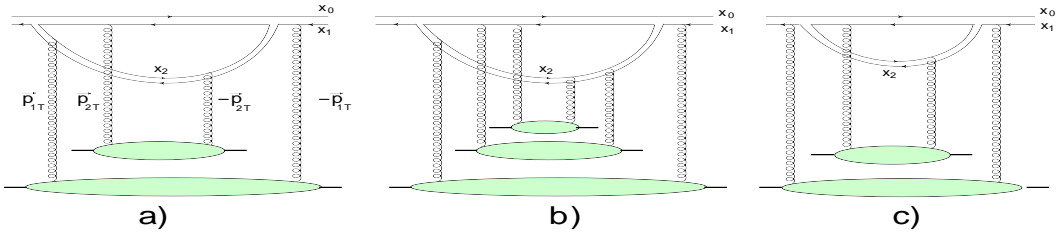} 
    \caption{ The interference diagram for the scattering of one dipole 
with two nucleons. The blocks show the amplitude : the exchange of two
 BFKL Pomerons(\fig{amn2}-a) and the contribution of the shadowing
 corrections to this diagram (\fig{amn2}-b).\fig{amn2}-c shows the
 diagrams that corresponds to the square of the amplitude. The blobs
 show the amplitudes of the gluon-nucleon interactions. }
\label{amn2}
  \end{figure}

The amplitude ${\cal N}^{(2)}\Lb Y, Y,\xi\Rb$ in the coordinate
 representation  is  intimately related to the scattering
 amplitude of the dipole of  size $x_{01}$  (see \fig{amn2})
 with the nucleus which interacts with the target two or more
 times. In the Born approximation such an amplitude is equal to
\beq \label{BESC4} 
\tilde{N}^{(2)}_{\rm B.A.}\Lb Y, Y,\tilde{\xi}; \vec{b} \Rb\,\,=\,\,\int d^2 b \Lb N^{(1)}_{\rm B.A.}\Lb Y, x_{01}, b \Rb\Rb^2/x_{01}^4\,
\,=\,\,\Big( \,\int d^2 b \, N^{(1)}_{\rm B.A.}\Lb Y, x_{01}, b \Rb\Big)^2\Big{/}\Lb x^4_{01}\,\pi \,R^2_A\Rb
\propto \ln^2(1/x^2_{01})\,\,\propto\,\,\tilde{\xi}^2
\eeq
where $\tilde{\xi}\,=\,\ln\Lb 1/r^2\Rb$. In DLA it is sufficient 
to change $\xi$ to $\tilde{\xi}$ for transforming from momentum to
 coordinate representation. 

The energy evolution of this amplitude is given by the emission
 of the extra gluon (see \fig{amn2}-c) and replaces $\tilde{\xi}$ in
 \eq{BESC4} by the function of $\tilde{\xi}$. The contribution of the
 interference diagram is shown in \fig{amn2}-a. One can see that gluons
 with transverse momenta $\vec{p}_{2T}$ and $- \vec{p}_{2T}$ interacts
 with the dipole $x_{12}$.  However, gluons with momenta $\vec{p}_{1T}$
 and $- \vec{p}_{1T}$ interact with the dipoles of the different sizes:
   $x_{12}$ and $x_{01}$. In the DLA $x_{12} \gg x_{01}$. The interaction
 with these two gluons  can be written as follows:
\bea \label{BESC5} 
&&\int d^2   b  \,N^{(1)}\Lb Y, x_{12},x_{01}, b\Rb \,\,\propto\,\int d^2\,p_{1T}\,\Lb 1\,+\,e^{i \vec{p}_{1T}\cdot \vec{x}_{12}}\Rb\,\Lb 1\,-\,e^{i \vec{x}_{01}\cdot \vec{p}_{1T}}\Rb\,\frac{N_G\Lb Y, p_{1T}\Rb}{p^4_{1T}}\nn\\
&&\,\xrightarrow{ p_{1T}^2 \,x^2_{12} \ll 1;  p_{1T}^2 \,x^2_{01}\,\ll\,1,x^2_{12} \,\gg\,x^2_{01}}\,\int_{1/x^2_{12}} d^2 p_{1T} \,p^2_{1T}\,x^2_{01}\,\frac{N_G\Lb Y, p_{1T}\Rb}{p^4_{1T}}\,\,\to\,\,\frac{x^2_{01}}{x^2_{12}} \int d^2 b\, N^{(1)}\Lb Y, x_{12}, b \Rb
\eea

 Using the result of \eq{BESC5}, we have the following expression for 
the diagram of \fig{amn2}-a:
 \bea \label{BESC6}
&&\frac{\partial   \int d^2 b \tilde{N}^{(2)}\Lb \fig{amn2} - a;Y, Y,\tilde{\xi}; \vec{b} \Rb}{\partial  \bas Y}\,=\\
&&\frac{1}{x^4_{01}\,\pi \,R^2_A}\Bigg\{\int d^2 x_{12} \int \frac{x^2_{01}}{x^2_{12}\,x^2_{02}}\,\int d^2 b \, N^{(1)}\Lb Y, x_{12}, b \Rb\Bigg\} \frac{x^2_{01}}{x^2_{12}} \int d^2 b\, N^{(1)}\Lb Y, x_{12}, b\Rb\nn\\
&&\xrightarrow{\rm DLA}\,\,\frac{1}{R^2_A}\int_{x^2_{01}}\,\frac{d x^2_{12}}{x^2_{12}}  \Bigg(\frac{ \int d^2 b\, N^{(1)}\Lb Y, x_{12}, b\Rb}{x^2_{12}}\Bigg)^2\nn
\eea

From \eq{BESC6}  we can re-write \eq{BESC2} 
 in the coordinate representation in a   more transparent  form. 
 It has the structure
\bea \label{BESC4}
\frac{\partial \int d^2 b \,\tilde{N}^{(2)}\Lb Y, Y,\tilde{\xi}; \vec{b} \Rb}{\partial \,\tilde{\xi}} \,\,&=&\,\,\int d^2 b \Bigg( \int^{Y}_{Y_0}\!\!\!\! d Y' \,\,\tilde{N}^{(2)}\Lb Y' ,Y,\tilde{\xi},\,\bv\Rb \Bigg\{2\,\,-\,\,e^{- \tilde{\xi}}\,n^{(1)}\Lb Y', \tilde{\xi},\bv\Rb\Bigg\}\nn\\
&+&\,\, 
\int^{Y}_{Y_0} \!\!\!\! d Y'\,\,\tilde{N}^{(2)}\Lb Y, Y', \tilde{\xi},\bv\Rb\, \Bigg\{2\,\,-\,\,e^{- \tilde{\xi}}\,n^{(1)}\Lb Y', \tilde{\xi},\bv\Rb\Bigg\}\nn\\
&+&\,\,\frac{2}{N^2_c - 1}
\int^Y\!\!\!\!  d Y'\,\tilde{N}^{(2)}\Lb Y', Y',\tilde{\xi},\bv\Rb \Bigg\{2\,\,-\,\,e^{- \tilde{\xi}}\,n^{(1)}\Lb Y', \tilde{\xi},\bv\Rb\Bigg\}\Bigg)
 \eea  
where  $n^{(1)}\Lb Y', \tilde{\xi},\bv\Rb\,=\,N^{(1)}\Lb Y', \tilde{\xi},
\bv\Rb\Big{/}r^2$.  All factors in \eq{BESC4} can be clarified using the
 Born Approximation for $N^{(1)} \,\propto \bas \ln Q^2$. Indeed,
 $\bas^2 \Lb 1/Q^2\Rb \ln \Lb Q^2\Rb \xrightarrow{\mbox{\tiny coordinate
 representation}} \bas^2 r^2 \ln\Lb r^2\Rb$. The ratio $\bas^2 r^2
 \ln\Lb r^2\Rb\Big{/} \Lb \pi R^2_A\Rb\,=\,\,N^{(1)}\Lb Y', \tilde{\xi},
\bv\Rb\,= r^2 \,n^{(1)}\Lb Y', \tilde{\xi},\bv\Rb$.  \eq{BESC4}
 is the double log limit of the general  non-linear equation\cite{LETU}
 (see \eq{AMP4}, for the terms without Bose-Einstein enhancement). This
 general equation has the following form for $\tilde{N}^{(2)}\Lb Y, Y,
\tilde{\xi}; \vec{b} \Rb$.

\bea \label{BESC5}
\int d^2 b\,\tilde{N}^{(2)}\Lb Y, Y,  \rv,\bv\Rb\,\,&=&
\,\bas\,\int^Y d Y' \int d^2 r' \frac{r^2}{r'^2\,\Lb \rv - \rv'\Rb^2}\,\nn\\
&\times& \Bigg(  \Big\{ 2\,\tilde{N}^{(2)}\Lb Y', Y, \rv',\bv- \h(\rv_i - \rv')\Rb  \,\,-\,\,\tilde{N}^{(2)}\Lb Y', Y \rv,\bv\Rb\,\nn\\
&-&\,\,\,\,
\tilde{N}^{(2)}\Lb Y', Y,\rv' ,\rv_{i+1},\bv - \h(\rv_i - \rv')\Rb\,N^{(1)}\Lb Y',\rv -  \rv' , \bv - \h \rv'\Rb\Big\}\nn\\
&+&  \Big\{ 2\,\tilde{N}^{(2)}\Lb Y, Y', \rv',\bv- \h(\rv_i - \rv')\Rb  \,\,-\,\,\tilde{N}^{(2)}\Lb Y, Y', \rv,\bv\Rb\,\nn\\
&-&\,\,\,\,
\tilde{N}^{(2)}\Lb Y, Y',\rv' ,\rv_{i+1},\bv - \h(\rv_i - \rv')\Rb\,N^{(1)}\Lb Y',\rv -  \rv' , \bv - \h \rv'\Rb\Big\}\nn\\
&+& \frac{2}{N^2_c - 1}  \Big\{ 2\,\tilde{N}^{(2)}\Lb Y', Y', \rv',\bv- \h(\rv_i - \rv')\Rb  \,\,-\,\,\tilde{N}^{(2)}\Lb Y', Y', \rv,\bv\Rb\,\nn\\
&-&\,\,\,\,
\tilde{N}^{(2)}\Lb Y', Y',\rv' ,\rv_{i+1},\bv - \h(\rv_i - \rv')\Rb\,N^{(1)}\Lb Y',\rv -  \rv' , \bv - \h \rv'\Rb\Big\}\Bigg)
 \eea
 Each term of this equation for $N^{(1)} = 1$ has
 the same form as \eq{BESC1}, including the term proportional to 
$\bas/(N^2-1)$ and, therefore, the asymptotic solution gives
 $\tilde{N}^{(2)}\Lb Y, Y,  \rv,\bv\Rb\,\to\,{\rm Const}$. In
 other words, the fast increase of the double gluon densities,
  which is generated by the Bose-Eintstein enhancement, turned
 out to be damped by the screening. It happens in spite of the
 power - like  increase of this amplitude  with energy in the
 linear evolution,  which is faster, than the increase for the elastic 
amplitude
 of dipole-target interaction.

  \section{Conclusions}
In this paper we discussed several facets of the energy
 evolution of the $J/\psi$  inelastic production cross
 section in deep inelastic processes. We showed that
 this cross section  in the linear approximation, can
 be written  by means of  the double gluon density, for which we can
 use the evolution equation in the BFKL kinematic region. This
 equation has been proposed previously in Refs.\cite{MUPA,PESCH,LELU}
(see Ref.\cite{DOS}   for a review of the equations in the DGLAP 
kinematic region).
  We completed this study by suggesting an equation for the
 generating functional which allows us to calculate multi-gluon
 density, that can be generated in the process of $J/\psi$ production.

We include in the evolution equation the Bose-Einstein enhancement
 which occurs at all values of energy (for example Ref.\cite{BEC4}).
We did this in the simplest case: DGLAP evolution in DLA approximation
 of perturbative QCD. Indeed, the particular form of the double gluon
 distribution function on which   the cross section of $J/\psi$
 production depends:
 $\int d^2 Q_T \,D^{(2)}\Lb x_1, \vec{p}_{1,T} + \h \vec{Q}_T;
 x_2, \vec{p}_{2,T} - \h \vec{Q}_T\Rb$, has a DLA limit. We found that
  the Bose-Einstein enhancement leads to  the  double gluon distribution
 function which increases faster that the product of two single gluon
 distribution function.  We discussed the value of the anomalous dimension
 as well as the solution of the resulting evolution equations. It turns out
 that the contribution of  the Bose-Einstein enhancement is rather small, 
and
 proportional to $\bas/\Lb N^2_c -\Rb)^2$, in accord with the estimates
 for twist four anomalous dimension\cite{BART1,BART2,LET2}. However, we
 believe that  understanding what happens to these contributions at
 ultra high energies, is a principal question for the effective theory,
 based on high energy QCD.

Bearing this in mind, we derived the evolution equation for the
 scattering amplitude of two dipoles with a nucleus, taking
 into account the shadowing corrections. We investigated the
 analytical solutions in two distinct kinematic regions: deep
 in the saturation region, and in the 
vicinity of the saturation scale.  Therefore, we have prepared the
 ground for building  saturation models to describe
 the $J/\psi$ production cross section.   Most data
 exists for the $J/\psi$ production, not in DIS but for the proton-nucleus
 interaction. As has been demonstrated in Refs.\cite{KLNT,DKLMT,KMV}
 this process is closely related to the production of $J/\psi$ in DIS.
 Hence, we need to re-write our formulae, since the
 distribution over transverse momentum of the  produced $J/\psi$ was
 measured,  which has not been discussed in this paper. We  plan
 to discuss this process elsewhere.

The suggested non-linear evolution equation is a direct generalization
 of the BK equation, and has to be solved  using the formulae of
 Refs.\cite{KLNT,DKLMT}, at the initial energy ($Y = Y_0$).
 The initial conditions depend on the saturation scale $Q_s\Lb Y=Y_0,b\Rb$.
  We need to compare the solution to the non-linear equation   
 in
 quasi-classical approach, in which we replace $Q_S\Lb Y=Y_0,b\Rb$ by 
$Q_s\Lb Y,b\Rb$  at arbitrary values of  rapidities $Y$.
  Such substitution looks  attractive from the physical 
point of 
view, since it takes into account the fact  that the main properties of 
the dense 
system of patrons depend on the saturation scale.  However,  we do not
 observe a small parameter that allows us to do this.
Consequently, it is instructive to compare the solution of the non-linear
 equation with this approximation, with the goal to find a new small 
parameter.

 Our last  result is that the shadowing corrections in the elastic
 amplitude, generate the survival probability that suppressed the energy
 behavior of the double gluon densities with the Bose-Einstein enhancement.
 We found that the cross section reaches a constant at ultra high energies.
 More precisely, we found that a constant is the asymptotic solution for
 the non-linear equation. Nevertheless, this constant could be equal
 to zero. In this case the cross section decreases as
 $\exp\Lb - \frac{z_1 + z_2)^2}{8 \kappa}\Rb$ being the 
solution of \eq{BESC1}.
If quasi-classical approach reflects the main property of
 the solution to the non-linear equation, we can expect 
from \eq{IC4} that, indeed, the cross section vanishes at high energies. 

In the paper we used the BFKL Pomeron calculus which is
  not as general as the CGC approach, being  equivalent
 to this approach only, in the limited range of rapidity. On the
 other hand, the  BFKL Pomeron calculus has two advantages:
  the possibility to treat on the same footing  elastic
 (diffractive)  and inelastic processes, and to   consider
  processes with different multiplicities.

  \section{Acknowledgements}
   We thank our colleagues at Tel Aviv university and UTFSM for
 encouraging discussions. Our special thanks go to  
 Alex Kovner,  Misha Lublinsky  and Marat Siddikov   for 
 discussions on the
 role of the Bose - Einstein correlation in the CGC  effective theory.
 
  This research was supported by the BSF grant   2012124, by 
   Proyecto Basal FB 0821(Chile),  Fondecyt (Chile) grant  
 1180118 and by   CONICYT grant PIA ACT1406.  
 
 ~    

\end{document}